\documentclass[12pt,a4paper,english]{JHEP3}

\usepackage[latin1]{inputenc}
\usepackage{amsmath,cite,graphicx}
\usepackage{babel}

\newcommand{\newc}{\newcommand}
\newc{\GeV}{\text{ GeV}}
\newc{\AS}{A(b)\sigma}
\newc{\Sinc}{\sigma^{\rm inc}}
\newc{\ASinc}{A(b)\cdot\sigma^{\rm inc}}
\newc{\db}{\mathrm{d}^2b \ }
\newc{\dpt}{\mathrm{d}p_T^2 \ }
\newc{\avgN}{\langle n(b, s) \rangle}

\newc{\ptmin}{p_T^{\rm min}}
\newc{\Pcal}{\mathcal{P}}
\newc{\UE}{Underlying Event}
\newc{\ue}{underlying event}
\newc{\Ue}{Underlying event}

\newc{\hw}{{\sf{Herwig++}}}
\newc{\vect}[1]{{\bf #1}}
\newc{\HWPP}{\textsf{Herwig++}}

\newc{\HWPPClass}[1]{\mbox{\href{http://projects.hepforge.org/herwig/doxygen/classHerwig_1_1#1.html}{\textsf{#1}}}}
\newc{\ThePEGClass}[1]{\mbox{\href{http://projects.hepforge.org/thepeg/doxygen/classThePEG_1_1#1.html}{\textsf{#1}}}}
\newc{\HWPPParameter}[2]{\mbox{\href{http://projects.hepforge.org/herwig/doxygen/#1Interfaces.html\##2}{{\bf #2}}}}

\renewcommand{\exp}[1]{\: e^{#1} \:}

\graphicspath{{./pics/}} 

\title{Simulation of multiple partonic interactions in Herwig++}

\author{Manuel B\"ahr, Stefan Gieseke\\ 
  Institut f\"ur Theoretische Physik\\
  Universit\"at Karlsruhe, 76128 Karlsruhe, Germany}
\author{Michael H. Seymour\\
  School of Physics and Astronomy, University of Manchester; and\\
  Physics Department, CERN, CH-1211 Geneva 23, Switzerland}

\abstract{In this paper we describe a new model of multiple partonic
interactions that has been implemented in \hw{}. Tuning its two free
parameters we find a good description of CDF \ue{} data. We show
extrapolations to the LHC.}
\keywords{Hadronic Colliders, QCD, Jets, Phenomenological Models, Underlying Event}
\preprint{CERN-PH-TH/2008-055\\KA-TP-08-2008}

\begin{document}

\section{Introduction\label{sec:Intro}}

With the advent of the Large Hadron Collider (LHC) in the near future it
will become increasingly important to gain a detailed understanding of
all sources of hadronic activity in a high energy scattering event.  An
important source of additional soft jets will be the presence of the
\ue{}.  From the experimental point of view, the \ue{} contains all
activity in a hadronic collision that is not related to the signal
particles from the hard process, e.g.\ leptons or missing transverse
energy.  The additional particles may result from the initial state
radiation of additional gluons or from additional hard (or soft)
scatters that occur during the same hadron--hadron collision.
Jet measurements are particularly sensitive to the \ue{} because,
although a jet's energy is dominated by the primary hard parton that
initiated it, jet algorithms inevitably gather together all other energy
deposits in its vicinity, giving an important correction to its energy
and internal structure.

In standard Monte Carlo event generators, like \textsf{Herwig(++)}
\cite{Corcella:2000bw,Gieseke:2003hm,Gieseke:2006ga,Bahr:2007ni,Bahr:2008pv,Bahr:2008tx}
, \textsf{PYTHIA} \cite{Sjostrand:2006za,Sjostrand:2007gs} or
\textsf{SHERPA} \cite{Gleisberg:2003xi}, additional gluons from initial
and final state radiation are generated with the help of parton
shower algorithms, possibly supplemented by multijet matrix
elements\cite{Catani:2001cc,Alwall:2007fs}.
Therefore, we tend to attribute these to the hard
process rather than to the \ue{}.  On top of that, the \ue{} is
simulated as some additional hadronic activity.  The simplest way to do
so is the so--called UA5 model \cite{Alner:1986is}, which has been the
default \ue{} model in \hw{} for a long time.  Here, additional (soft)
hadronic activity is generated as a number of additional clusters are
generated flat in rapidity with an exponential transverse momentum
distribution.  See \cite{Bahr:2008pv} for more details.  These clusters
eventually give the required additional activity of soft hadrons.

Another variant, which has been far more successful in the description of
recent collider data, was formulated as a sequence of more-or-less
independent parton interactions.  In contrast to the UA5 model this
model is capable of describing the jet--like structure of the \ue{}. In
its initial formulation \cite{Sjostrand:1987su} there were no parton
showers invoked.  Later variants of this model also contain full parton
showers \cite{Butterworth:1996zw,JimmyManual}.  The additional scatters
in these models are always modelled as simple QCD $2\to 2$ scattering as
long as the scattering contains a hard jet of at least a few GeV.  Soft,
more forward scattering may also be modelled but requires a unified
description of perturbative and non--perturbative scattering, as in the
dual parton model \cite{Capella:1978ig,Capella:1980fv,Capella:1981xr},
which had been implemented into the event generator \textsf{PHOJET}
\cite{Engel:1994vs}.  Another model is the simple extrapolation of the
transverse momentum distribution of hard jets in QCD processes down to
zero $p_T$ \cite{Borozan:2002fk}.  Such a modelling of soft interactions
will also allow us to describe minimum bias events.  These are dominated
by soft, forward scatterings and diffractive production of particles
during the hadron--hadron scattering event.

Experimentally, there has been strong evidence for the presence of
multiple partonic interactions already at the CERN ISR through the
measurement of a momentum imbalance in multijet events
\cite{Akesson:1986iv}.  The idea for this measurement is that multiple
pairs of jets, two in this case, will appear to be balanced in
transverse momentum if they have been created in different
back--to--back events rather than a single multijet event. Similar
observations of double parton scattering \cite{Pumplin:1997ix} have been
made at the Tevatron \cite{Abe:1993rv,Abazov:2002mr}.  Nowadays, the
clearest observation has been made in $\gamma$+3 jet events at CDF
\cite{Abe:1997xk}.  In addition to this clear evidence for the presence
of multiple interactions in hadronic collisions, the only sensible
description of the final state of such events can be made with detailed
Monte Carlo modelling, based on this ansatz.  The most detailed
measurements of the properties of the underlying event as well as their
implications for Monte Carlo models are described in
\cite{Affolder:2001xt,Acosta:2004wqa}.

Understanding minimum bias interactions and the \ue{} are very important
for many aspects of LHC physics.
Particularly in high luminosity runs, every triggered hard event in one
bunch crossing will be accompanied by additional interactions among
other protons from the same bunch.  These are predominantly minimum bias
interactions and will give some additional activity in the detectors.
There are already detailed plans for the measurement of the \ue{} in
ATLAS \cite{Moraes:2007rq} and CMS \cite{Fano:2007zz,Acosta:2006bp}.
The presence of the \ue{} is important whenever measurements at the LHC
will be based on the measurements of the properties of jets, like e.g.\
their energy.  The determination of the so--called jet energy scale is
known to be improved when a reasonable modelling of the \ue{} is
included in the analysis.  A good example for this is the measurement of
the top mass \cite{Borjanovic:2004ce}.  Implications for the central jet
veto in vector boson fusion processes have been addressed in detail in
\cite{ChristophThesis}.

In this paper we want to focus on the description of the hard component
of the \ue{}, which stems from additional hard scatters within the same
proton.  Not only does this model give us a simple unitarization of the
hard cross section, it also allows to give a good description of the
additional substructure of the \ue{}s.  It turns out that most activity
in the \ue{} can be understood much better in terms of hard minijets.
We therefore adopt this model, based on the model \textsf{JIMMY}
\cite{Butterworth:1996zw,JimmyManual}, also for our new event generator
\hw{} \cite{Bahr:2008pv}.  We will describe the basic implementation of
the model and its parameters and study some important implications for
jet final states.  Thus far, we do not consider a description beyond
multiple hard interactions.  An extension of our model towards softer
interactions along the lines suggested in \cite{Borozan:2002fk} is
planned and will also allow us to describe minimum bias interactions. 

Improvements of the underlying event description have also been
implemented in other event generators.  A completely new formulation of
the interleaving of underlying hard scatterings with the parton shower
has been introduced with the latest versions of \textsf{PYTHIA}
\cite{Sjostrand:2004pf,Sjostrand:2004ef,Sjostrand:2006za,Sjostrand:2007gs}.
A model very similar to the multiple interaction model in
\textsf{PYTHIA} has been implemented in \textsf{SHERPA}
\cite{Alekhin:2005dx}.  A new approach, based on $k_T$--factorization
\cite{Catani:1990eg,Collins:1991ty,Levin:1991ya} has been introduced and
studied in \cite{Hoche:2007hg}.  An important issue, which
has been addressed in \cite{Sjostrand:2004pf} is the relation between
the charged particle multiplicity and the average transverse momentum in
the underlying event.  The relation between these observables in the
transverse region of jet events may point us towards the right colour
correlations of the different hard scatters \cite{Bartels:2005wa}.  We
want to point out that the organization of colour lines adopted in our
model differs significantly from that in \textsf{PYTHIA}.  In this paper
we would like to focus on the details of the implementation, validation
and tuning on Tevatron data and some predictions for the LHC.

This paper is organized as follows: In Sect.~\ref{sec:details} we briefly
review the theoretical motivation for multiple interactions and describe
all details that are relevant for our Monte Carlo implementation.  In
Sect.~\ref{sec:Results} we discuss the parameters of our model and perform
a fit to current Tevatron data.  Taking this as a starting point,
we make predictions for the most important final state observables at
the LHC. Furthermore, we discuss implications of the intrinsic
uncertainties of parton distribution functions for the \ue{}
observables. In Sect.~\ref{sec:Conclusions} we draw some conclusions and
give an outlook to future work. Some more model details will be described
in Appendices.

\section{Details}
\label{sec:details}

The starting point for thinking about multiple interactions is the
observation that the cross section for QCD jet production may exceed the
total $pp$ or $p\bar p$ cross section already at an intermediate energy
range and eventually violates unitarity.  For example, for QCD jet
production with a minimum $p_T$ of 2 GeV this already happens at
$\sqrt{s} \sim 1$~TeV. This $p_T$ cutoff should however be large enough
to ensure that we can calculate the cross section using pQCD.  The
reason for the rapid increase of the cross section turns out to be the
strong rise of the proton structure function at small $x$, since the $x$
values probed decrease with increasing centre of mass energy. This
proliferation of low $x$ partons may lead to a non-negligible
probability of having more than one partonic scattering in the same
hadronic collision. This is not in contradiction with the definition of
the standard parton distribution function as the \emph{inclusive}
distribution of a parton in a hadron, with all other partonic
interactions summed and integrated out. It does, however, signal the
onset of a regime in which the simple interpretation of the pQCD
calculation as describing the only partonic scattering must be
unitarized by additional scatters.

In principle, predicting the rate of multi-parton scattering processes
requires multi-parton distribution functions, about which we have almost
no experimental information.  However, the fact that the standard parton
distribution functions describe the inclusive distribution gives
a powerful constraint, which we can use to construct a simple model.

\subsection{Eikonal model}

The eikonal model introduced in
Refs.~\cite{Durand:1987yv,Durand:1988ax,Butterworth:1996zw} derives from
the assumption that at fixed impact parameter $b$, partons undergo
independent scatters with mean number
\begin{equation}\label{eq:average1}
  \begin{split}
    \langle n(b=|\vect{b}|, s)\rangle = & \int \mathrm{d}^2\vect{b}^{\prime}
    \int_{{\ptmin}^2} \dpt \sum_{ij} \frac{1}{1+\delta_{ij}} \
    \frac{\mathrm{d}\hat\sigma_{ij}(x_1\sqrt{s}, x_2\sqrt{s}, p_T^2)}
         {\mathrm{d}p_T^2} \\
    &\otimes G_{i/h_1}(x_1, \vect{b}-\vect{b}^{\prime},\mu^2)
    \otimes  G_{j/h_2}(x_2, \vect{b}^{\prime},\mu^2) \, ,
  \end{split}
\end{equation}
where d$\hat\sigma$ is the differential partonic cross section for QCD
2$\to$2 scattering and $G(x, \vect{b}, \mu)$ are parton densities
representing the average number of partons with a given momentum
fraction $x$ and transverse coordinate $\vect{b}$. $\otimes$ denotes the
convolution integrals in longitudinal momentum fractions $x_1, x_2$.  By
further assuming a factorization of the $x$ and $\vect{b}$ dependence in
$G$, namely
\begin{equation}
  G(x, \vect{b}, \mu^2) = f(x, \mu^2) \cdot S(\vect{b}) \, ,
\end{equation}
where $f(x, \mu^2)$ is the conventional parton distribution,
Eq.~(\ref{eq:average1}) can be written as
\begin{align}
  \avgN =& \Sinc(s;\ptmin) \cdot 
  \int \mathrm{d}^2\vect{b}^{\prime} \ S_{h_1}(\vect{b}^\prime) \
  S_{h_2}(\vect{b} - \vect{b}^\prime) \nonumber\\
  =& \ASinc(s;\ptmin) \, .\label{eq:average2}
\end{align}
In Eq.~(\ref{eq:average2}) $\Sinc$ denotes the inclusive cross section
to produce a pair of jets (partons) with $p_T > \ptmin$ and is given by
the standard perturbative calculation, whereas $A(b)$
describes the overlap of the partons in the colliding hadrons.
We model the impact parameter dependence of partons in a hadron,
$S(\vect{b})$, by the
electromagnetic form factor,
\begin{equation}\label{eq:formfactors}
  S_{\bar p}(\vect{b}) = S_p(\vect{b}) = \int \frac{\mathrm{d}^2\vect{k}}{2\pi} \
  \frac{\exp{i \vect{k} \cdot \vect{b}}}{(1+\vect{k}^2/\mu^2)^2} \, ,
\end{equation}
where $\mu$ is the inverse hadron radius. This leads to
\begin{equation}\label{eq:overlap2}
  A(b) = \frac{\mu^2}{96 \pi} (\mu b)^3 K_3(\mu b) \, ,
\end{equation}
where $K_3(x)$ is the modified Bessel function of the third kind. We do not
fix $\mu$ at the value determined from elastic $ep$ scattering, but
rather treat it as a free parameter, because the spatial parton
distribution is assumed to be similar to the distribution of charge, but
not necessarily identical.

The assumption that different scatters are uncorrelated leads to the
Poissonian distribution for the number of scatters, $k$, at fixed impact
parameter,
\begin{equation}
  \Pcal_k(\langle n \rangle) = \frac{{\langle n \rangle}^k}{k!} \exp{-{\langle n \rangle}} \, .
\end{equation}

The cross section for having exactly $n$ scatters with individual cross
section $\Sinc$, using this assumption is
\begin{equation}
  \sigma_n(\Sinc) = \int \db \ \Pcal_n(\ASinc) = \int \db \
  \frac{(\ASinc)^n}{n!} \exp{-\ASinc} \, .
\end{equation}
The probability of having $n$ scatters in an event, given that there is
at least one, is then
\begin{equation}\label{eq:prob0}
  \mathrm{P}_{n \geq 1}(\Sinc) = \frac{\int \db \ \Pcal_n(\ASinc)}{\int
    \db \sum_{k=1}^{\infty} \ \Pcal_k(\ASinc) } =
  \frac{\sigma_n(\Sinc)}{\sigma_{hard}(\Sinc)} \, . 
\end{equation}
Equation~(\ref{eq:prob0}) is used as the basis of the multi-parton
scattering generator for events in which the hard process is identical
to the one used in the \ue, i.e.~QCD $2\to2$ scattering.
For distinct scattering types a modification is used, as described in
the next section.

\subsubsection{Different scattering types}

Following the assumption of independent additional scatterings the cross
section for two distinct scattering types $a$ and $b$ with the respective
multiplicities $k$ and $m$ can be written as
\begin{align}\label{eq:2scatterings}
  \sigma_{k,m}(\sigma_a, \sigma_b) &= \int \db \ \Pcal_k(\AS_a) \
  \Pcal_m(\AS_b) \nonumber\\
  &= \int \db \frac{(\AS_a)^k}{k!} \exp{-\AS_a} \ \frac{(\AS_b)^m}{m!}
  \exp{-\AS_b} \, .
\end{align}
For small signal cross sections $\sigma_b$, the exponential can be
approximated by unity. Using Eq.~(\ref{eq:2scatterings}) the probability of having
$k$ events of type $a$ in the presence of exactly one of type $b$ is
\begin{align*}
  \mathrm{P}_{k} &= \frac{\sigma_{k,1}}{\sum_{\ell=0}^{\infty} \sigma_{\ell,1}}
         \approx \frac{\int \db \ \Pcal_k(\AS_a) \cdot \AS_b}{\int \db \ \AS_b} \\
         &= \int \db \  \Pcal_k(\AS_a) \cdot A(b) \, .
\end{align*}
This can then be rewritten to avoid the extra factor $A(b)$ in the form,
\begin{align}\label{eq:prob1}
  P_{n=k+1} &\approx \frac{n}{\sigma_a} \int \db \ \Pcal_n(\AS_a) \, .
\end{align}
Here $n$ is the total number of scatters, i.e.\ there is one of type $b$
and $n-1$ of type~$a$.
It is worth noting that the fact that we have `triggered on' a process
with a small cross section leads to a bias in the $b$ distribution and
hence a higher multiplicity of additional scatters than in the pure
QCD $2\to2$ scattering case.

Equation~(\ref{eq:prob1}) can be used to describe \ue{} activity under
rare signal processes as well as jet production in the \ue{}
simulated under high $p_T$ jet production as signal process. In the latter
case the assumption of distinct scattering processes may not be fulfilled.
One can show that in
that case the $m$th scatter of type $a$ that is also of type $b$ should
be rejected
with probability $1/(m+1)$.

\subsection{Monte Carlo implementation}

The model introduced so far is entirely formulated at the parton level.
However, an event generator aims for a full description of the event at
the level of hadrons. This implies that the implementation of
multi-parton scattering must be properly connected to the parton shower
and hadronization models, a few details of which we discuss in the
following.  We give more technical details of the way in which the
multiple scattering is represented in the event record, and of how to
access the model parameters, in Appendices~\ref{appA} and \ref{appB}
respectively.

Event generation starts with the sampling of the hard process
according to its matrix element and the parton densities. After that the
parton shower evolves the final state partons from the scale of the
hard interaction down to a cutoff scale that is of the order
of the confinement scale, but large enough to ensure that we remain
within the perturbative regime. The
incoming partons are evolved backwards to 
higher values of $x$ and decreasing $\mu^2$. The initial- and final-state
parton showers in \HWPP\ are performed using the coherent branching algorithm
of Ref.~\cite{Gieseke:2003rz}, which is based on the original coherent shower
algorithm of Refs.~\cite{Webber:1983if,Marchesini:1984bm,Marchesini:1987cf}. 
After the initial-state shower has terminated, the incoming partons are
extracted out of the beam particles, a step that we describe in more
detail below.

Now the number of secondary interactions is sampled from the probability
distributions of Eq.~(\ref{eq:prob0}) or Eq.~(\ref{eq:prob1})
respectively. The chosen number of additional scatters is sampled
according to the standard QCD $2\to2$ matrix elements and the same
parton densities that were used for the hard process. That is, the
additional hard processes are generated exactly according to the
inclusive perturbative cross section, with no modification for the fact
that they are additional scatterings. This list of processes 
is then successively processed by the parton shower.
The partons involved in the additional hard scatters are also parton
showered.  As far as final state showering is concerned, this is
identical to a standard hard process.  For the initial state shower, we
use the standard evolution algorithm, but with modified parton
distribution functions, motivated by our model for extracting partons
out of the hadron, which we return to shortly.

If the backward evolution of a scattering leads to a violation of
four-momentum conservation, the scattering cannot be established. It is
therefore regenerated until the desired multiplicity has been
reached. If a requested scattering can never be generated without
leading to violation of momentum conservation, the program eventually
gives up, reducing the multiplicity of scatters.

After the parton shower, the quarks and gluons must be formed into the
observed hadrons. The colour preconfinement property of the angular-ordered
parton shower is used as the basis of the cluster model~\cite{Webber:1983if},
which is used in \HWPP\ to model the hadronization. 
The cluster model however necessarily expects (anti)quarks or
(anti)diquarks at the beginning of the hadronization. In the final state
this prerequisite is easily fulfilled by the gluon splitting mechanism:
all final-state gluons decay non-perturbatively to light
quark--antiquark pairs. In the case of an initial-state parton from an
incoming hadron, this necessitates a parton extraction model, which we
describe in the next section.

Finally all unstable particles must be decayed. \HWPP\ uses a
sophisticated model of hadronic decays as described in
Refs.~\cite{Grellscheid:2007tt,MesonDecays}.

\subsubsection{Parton extraction}

In the standard \HWPP\ treatment of a single hard scattering, the
prerequisite that the outgoing partons must be (anti)quarks or
(anti)diquarks is implemented by forcing the backward evolution to
terminate on a valence parton.  This then gives a diquark as the proton
remnant for example.  This diquark is colour-connected through the
colour connections of the valence quark either to a final-state parton
emitted during the corresponding initial-state parton shower or through
the hard process to a parton in one of the other jets in the event.  In
collisions other than $pp$, in events with little radiation, it can even
be connected right through the event to the other hadron remnant.

It is often the case that by the time the perturbative evolution has
terminated, the backward evolution has reached a valence parton, since
their PDFs dominate at high~$x$ and low scale.  When this is not the
case and the backward evolution has terminated on a gluon or sea quark,
one or two additional backward steps respectively are `forced', using
the standard backward evolution algorithm, but with all flavours except
the one necessary for the forced step, vetoed.

In the implementation of multiple interactions, we keep the treatment of
the first interaction untouched, i.e.\ it is exactly as just described.
This means that the valence structure of the hadron has already been
saturated, with one valence parton extracted and the remainder forming
the hadron remnant.  This does not therefore provide a structure that
can be iterated for subsequent scatters.  Instead, we modify the
backward evolution so that it terminates on a gluon.  We do this both
dynamically during the evolution and by a forced backward evolution step
if necessary.  During the backward evolution we use modified parton
distribution functions that are identical to the standard ones but with
the valence contributions subtracted out\footnote{They do not therefore
obey a momentum sum rule, but the algorithm is not sensitive to this
fact, since it only involves ratios of PDFs.  If one wanted to, one
could rescale all the modified PDFs by a common factor to regain the
momentum sum rule.  The results would be unchanged.}.  We stress that
this subtraction of valence contributions is the only modification we
make.  In particular, the distribution of gluons is identical to that in
the original hadron, leading to the possibility that the backward
evolution of multiple scatters can over-saturate the available energy,
which we deal with as already discussed above.

Once the backward evolution has terminated on a gluon, its colour
connections can therefore be inserted into those of the previous
remnant.  As a concrete example, for the second scattering in an event
with an incoming proton, the colour line of the gluon is connected to
the diquark proton remnant and the anticolour line of the gluon is
connected through the valence quark, to the outgoing parton that the
diquark was previously connected to.  This then gives a structure that
can be iterated an arbitrary number of times.  Since we do not order the
additional hard scatters, for example in transverse momentum, this is
equivalent to the colour connection model described as `random' in
\cite{Sjostrand:2004pf}.  The implementation of other colour connection
models as described there would be possible, and may be interesting work
for the future.

\FIGURE[t]{%
  \includegraphics[scale=0.7]{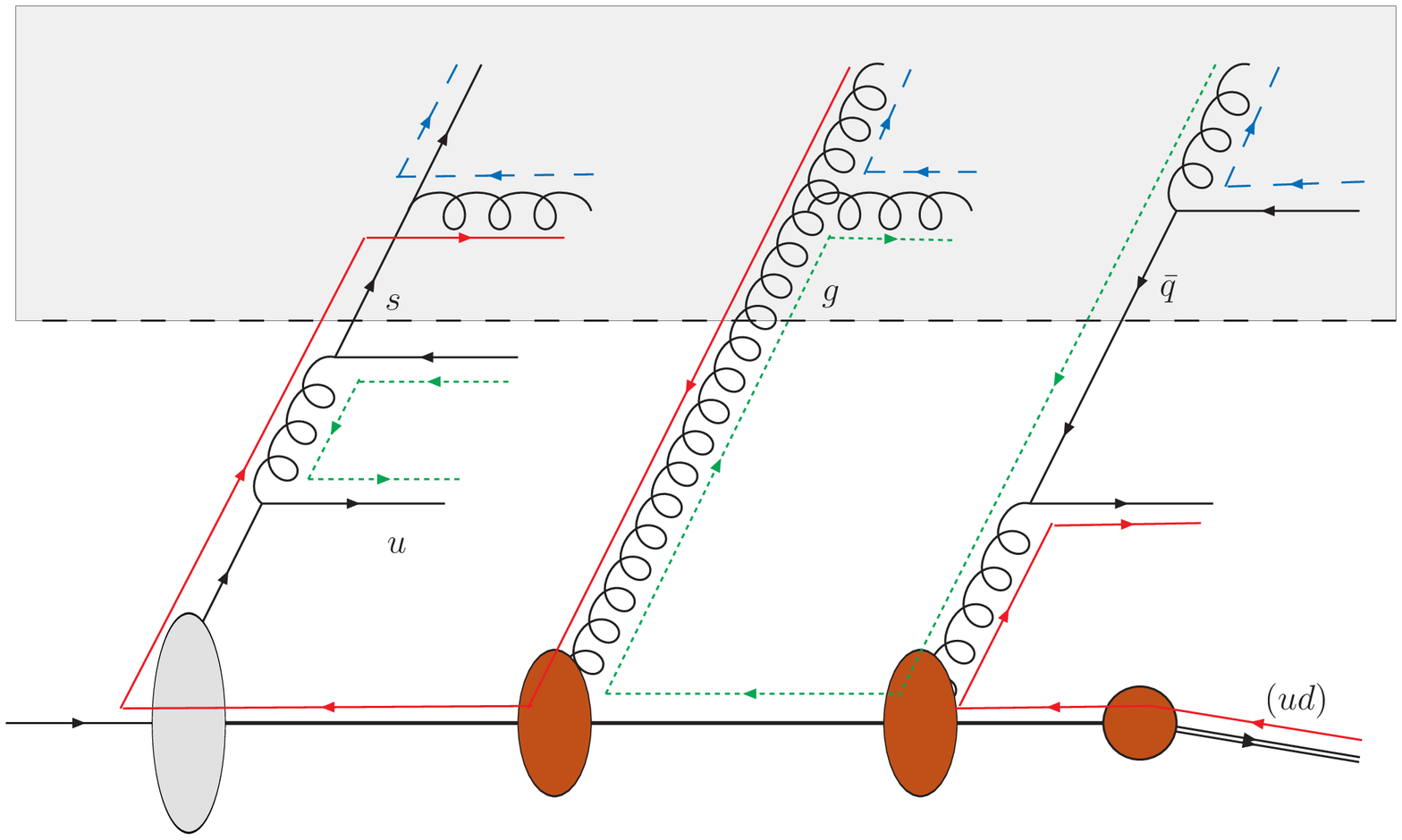}
  \caption{
    \label{fig:schema}
    Schema of how the forced splittings and colour connections are
    implemented.  Splittings in
    the shaded area stem from the hard scatters and the initial state
    parton shower.  The final splittings at the bottom are
    non--perturbative.}
} 
We illustrate this parton extraction model in more detail in
Fig.~\ref{fig:schema}.  In the upper part
of the figure, which is shaded, we can see the extracted partons after a
possible perturbative parton shower.  In the lower half of the figure,
additional forced splittings are carried out in order to guarantee a
certain flavour structure of the remnant.  The first extracted parton
will always be a valence quark while all additional hard scatters will
always end up on a gluon.  The colour structure is as just described,
with the gluon produced by each hard scatter inserted into the
colour--anticolour connection left by the previous one.

The way in which the structure of the hadron remnant is represented in
the event record is not quite the same as the way in which it is
generated, as described above.  The same event is shown in
Fig.~\ref{fig:schema2} as it would appear in the event record, as
described in Appendix~\ref{appB}.

\section{Results\label{sec:Results}}

We will now discuss several hadronic observables both for the Tevatron and
the LHC. In particular a comparison to CDF data \cite{Affolder:2001xt} is
performed. For that reason the non-standard jet algorithm used for the
data analysis has been implemented. Detector effects are solely taken
into account by simulating the 92\% track efficiency simply by ignoring
8\% of charged particles, chosen randomly. For the LHC the
prediction is compared to several other generators \cite{Alekhin:2005dx}.

\subsection{Tuning and Tevatron results}\label{sec:TVTtune}

\FIGURE[t]{
  \includegraphics[%
    width=0.48\textwidth,keepaspectratio]{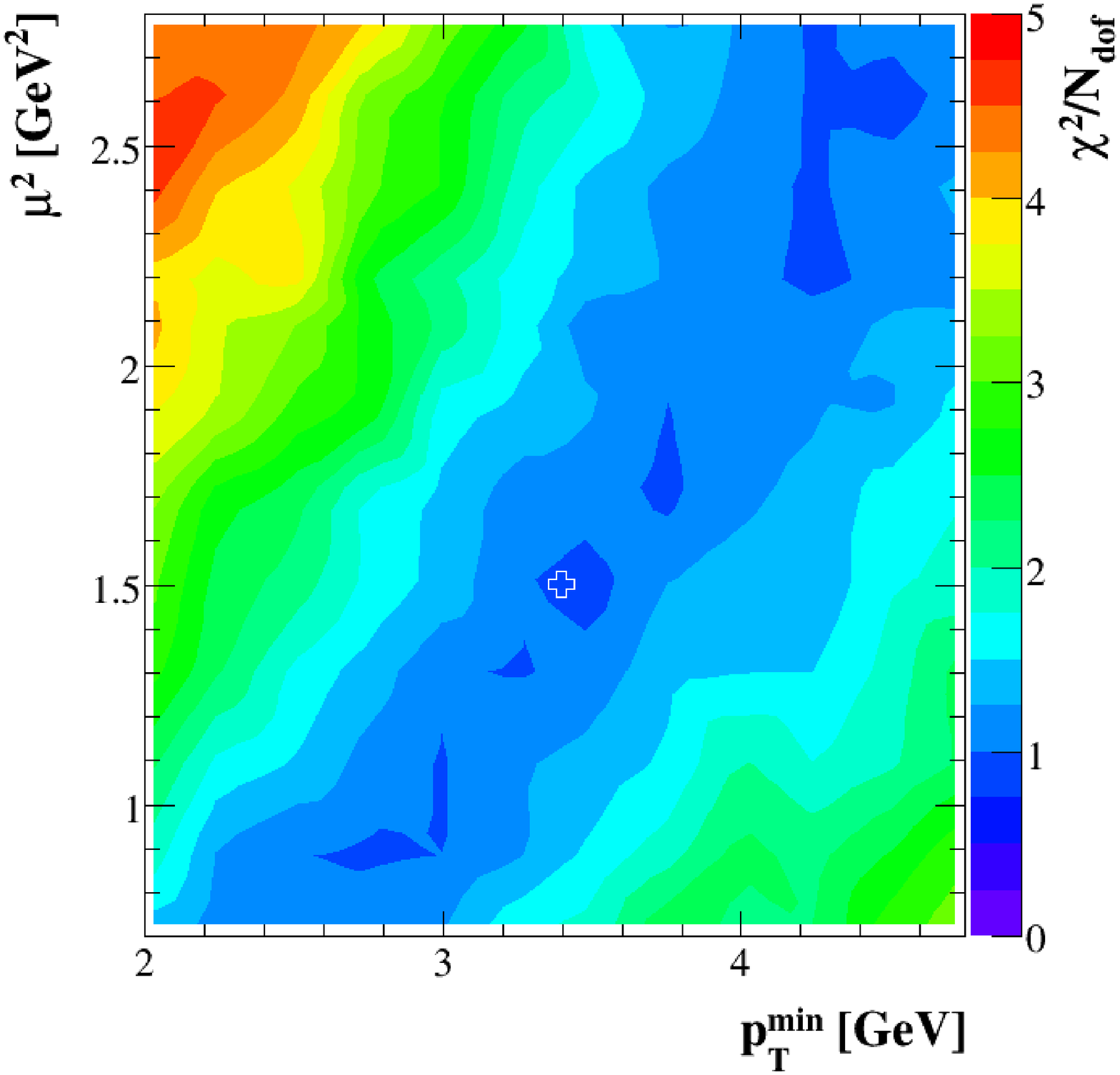}
  \includegraphics[%
    width=0.48\textwidth,keepaspectratio]{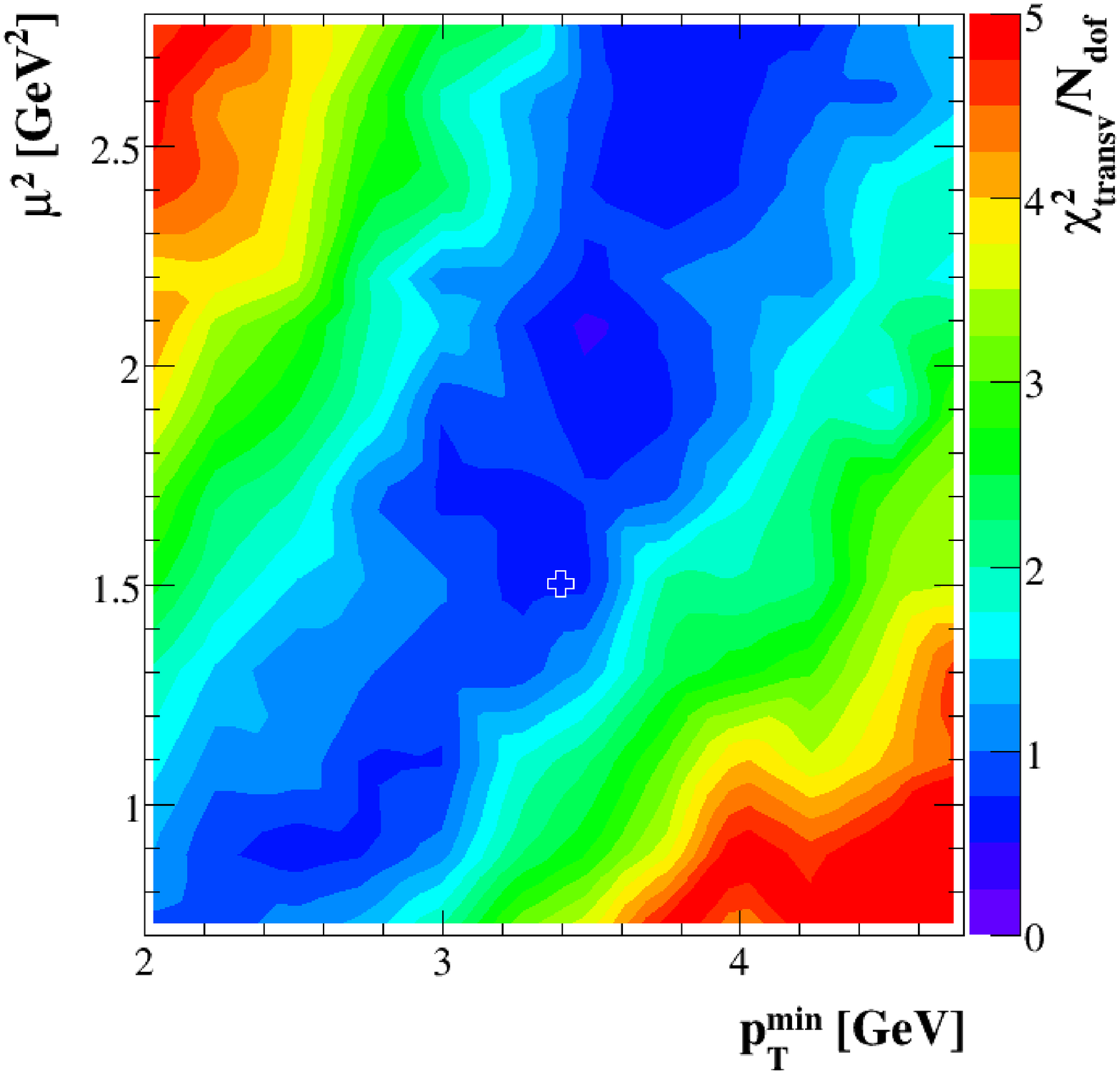}
  \caption{
    \label{fig:scans}
    Contour plots for the $\chi^2$ per degree of freedom of all
    discussed observables (left) and only the ones from the transverse
    region (right). The cross indicates the location of our preferred
    tune.}
}

We have performed a tune of the model by calculating the total $\chi^2$
against the data from Ref.~\cite{Affolder:2001xt}. For this analysis each
event is partitioned into three parts, the \textbf{towards, away} and
\textbf{transverse} regions. These regions are equal in size in $\eta -
\phi$ space and classify where particles are located in this space with
respect to the hardest jet in the event. We compare our predictions
to data for the average number of charged particles and for the scalar
$p_T$ sum in each of these regions. As we are aiming primarily at a good
description of
the \ue{} in high $p_T$ events, we used jet production with a
minimal transverse momentum of 15 GeV as the signal process. Because of
that we only use data for the region of leading jet transverse momentum
above 20~GeV. We added an additional systematic error in quadrature for the
lowest $p_T$ bins as described in Appendix~\ref{app:syserr}.

The parameter space for this tune is two dimensional and consists of the
$p_T$ cutoff $\ptmin$ and the inverse hadron radius squared, $\mu^2$,
entering $A(b)$ in Eq.~(\ref{eq:formfactors}). In Fig.~\ref{fig:scans} we
show the contour plots for all six observables and for the observables
from the transverse region respectively.
We have used the MRST~2001 LO\cite{Martin:2001es} PDFs built in to \hw{}
for this plot, and discuss the PDF-dependence in the next section.  For
these, and all subsequent plots, we use \HWPP\ version 2.2.1, with all
parameters at their default values except the two we are tuning and, in
the next section, the PDF choice.

The description of the Tevatron data is truly satisfactory for the
entire range of considered values of $\ptmin$. For each point on the
$x$-axis we can find a point on the $y$-axis to give a reasonable
fit. Nevertheless an optimum can be found between 3 and 4 GeV. The strong and
constant correlation between $\ptmin$ and $\mu^2$ is due to the fact
that a smaller hadron radius will always balance against a larger $p_T$
cutoff as far as the \ue{} activity is concerned.

As a default tune we use $\ptmin = 3.4 \GeV$ and $\mu^2 = 1.5 \GeV^2$.
Figure~\ref{fig:transv} shows the result of this parameter choice for
the transverse region. The towards region is shown in
Fig.~\ref{fig:towards} as well as the away region in
Fig.~\ref{fig:away}. For these plots we used 10 million events in contrast to
1 million for each point in Fig.~\ref{fig:scans}, which is the reason
for the slight
differences in the corresponding $\chi^2$ values.

It is clear from these figures that event generation without any model
for the \ue{} is not capable of describing the data. In
particular, in the transverse region, which receives the least
contribution of the two jets from the matrix element, the results are a
factor of two below the data.

Although our default multi-parton interaction (MPI) model gives a good
overall description of the data, we see a slight trend to produce too
much multiplicity in all the regions, most noticeably in the towards
region, and too little $p_T^{sum}$ in all the regions, most noticeably
in the away region.  This corresponds to having a slightly too soft
spectrum of individual particles and has also been observed in attempts
to fit the fortran \textsf{HERWIG+Jimmy} model, the forerunner of ours,
to the data of \cite{Affolder:2001xt}. We note that in the
towards region, which is dominated by the primary jet, \hw{} without MPI
is already close to the data, leaving very little room for MPI effects.
Almost any model of the underlying event will produce more than enough
multiplicity here and overshoot the data.  The same is true to a lesser
extent in the away region.  In the process of $\chi^2$ minimization,
there is therefore a slight pressure to suppress the \ue{} effect, which
results in the slight undershooting of the $p_T^{sum}$ predictions.  The
same effect is true even more weakly in the transverse region, where
one would say that the description is very good, but there is a slight
trend to be above the data for $N_{chg}$ and below it for $p_T^{sum}$.
Since the effect is strongest for the regions dominated by the primary
jets, we conclude that this is a general \textsf{Herwig} issue not
specifically related to the MPI model.  In any case, it is clear that it
vastly improves the description of data relative to the no-MPI model.

We want to stress that the data from the experimental analysis are
uncorrected. We already obtain a total $\chi^2$ per degree of
freedom very close to unity even with an over-simplified implementation
of the reconstruction efficiency but a more precise examination would
have to take detector effects into account in a more complete manner.

\subsection{PDF uncertainties}

For precision studies it is important to quantify the extent to which
hard scattering cross sections are uncertain due to uncertainties in the
PDFs.  As we have already mentioned, jet cross sections are particularly
sensitive to the amount of \ue{} activity, which introduces an
additional dependence on the PDF in our model.  In particular, it relies
on the partonic scattering cross sections down to small transverse
momenta, which probe momentum fractions as small as $x\sim10^{-7}$ at
the LHC and $x\sim10^{-6}$ at the Tevatron, where the PDFs are only
indirectly constrained by data.  One will have measured the amount of
\ue{} activity at the LHC by the time precision measurements are being
made, so one might think that the size of the \ue{} correction will be
known.  However, in practice, jet cross section corrections depend
significantly on rare fluctuations and correlations in the \ue{}, so the
correction must be represented by a model tuned to data, rather than by
a single number measured from data.  This will therefore entail in
principle a retuning of the parameters of the \ue{} model for each new
PDF.  This would make the quantification of PDF errors on a given jet
cross section, or of extracting a new PDF set from jet data, much more
complicated than a simple reweighting of the hard scattering cross
section.

In this section we explore the extent to which this effect is important,
by studying how the predictions with fixed parameters vary as one varies
the PDF.  We do this by comparing the central values of two different
PDF sets (MRST and CTEQ) and also using the quantification of the
uncertainties within one of them (CTEQ).  Similar issues were also
discussed in Ref.~\cite{Gieseke:2004tc} for the uncertainty in parton
shower corrections, which were found to be relatively small.

The results in Figs.~\ref{fig:transv}--\ref{fig:away} show the
predictions of our model with MRST~2001\cite{Martin:2001es} and
CTEQ6L\cite{Pumplin:2002vw} PDFs with the parameters fixed to the values
obtained from our fit with the MRST PDFs.  We see that the difference in
the amount of underlying event activity, quantified by the results in
the transverse region between 30 and 40~GeV as an example, is some 10\%
higher
with CTEQ6L than with MRST.

To quantify the effect of the uncertainties within a given PDF set, we
have used the error sets provided with the CTEQ6 family, and the formula
\begin{equation}
  \Delta X = \frac{1}{2} \ 
  \left( 
    \sum_{i=1}^{N_p} \left[ X(S_i^+) - X(S_i^-) \right]^2  
  \right)^{1/2} \ ,
\end{equation}
from Ref.~\cite{Pumplin:2002vw}. Here $X$ is the observable of
interest and $X(S_i^\pm)$ are the predictions for $X$ based on the PDF
sets $S_i^\pm$ from the eigenvector basis. Doing this na\"\i vely, we
found that the statistical error on independent runs with each PDF set
was greater than the variation between the sets. To try to overcome this
obstacle, we have studied the relative PDF uncertainty, i.e.\ $\Delta X
/ X(S_0)$, as a function of the number of points used for each
$X(S_i^\pm)$.
\FIGURE[t]{
  \includegraphics[%
    width=.48\textwidth,keepaspectratio]{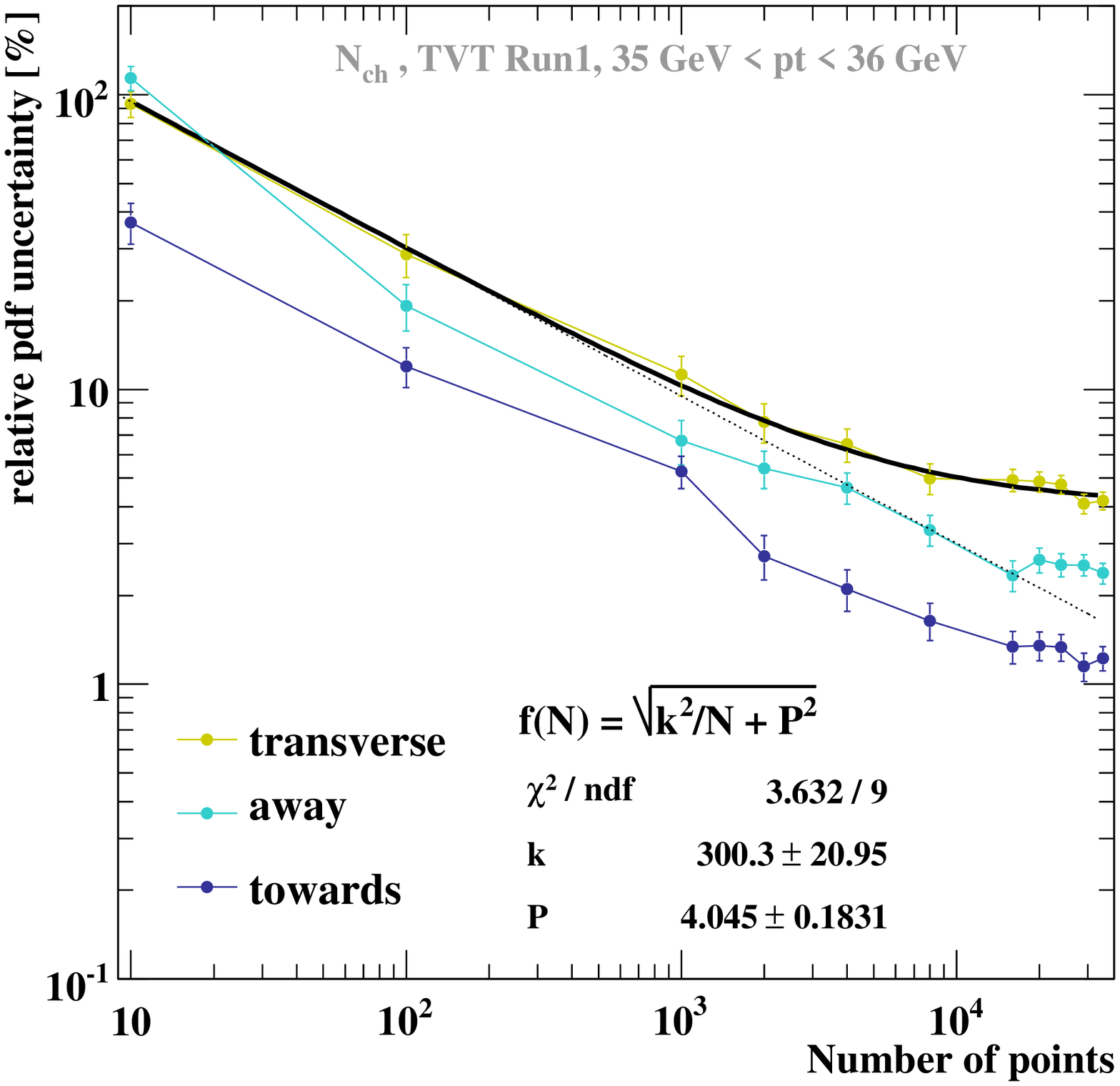}
  \includegraphics[%
    width=.48\textwidth,keepaspectratio]{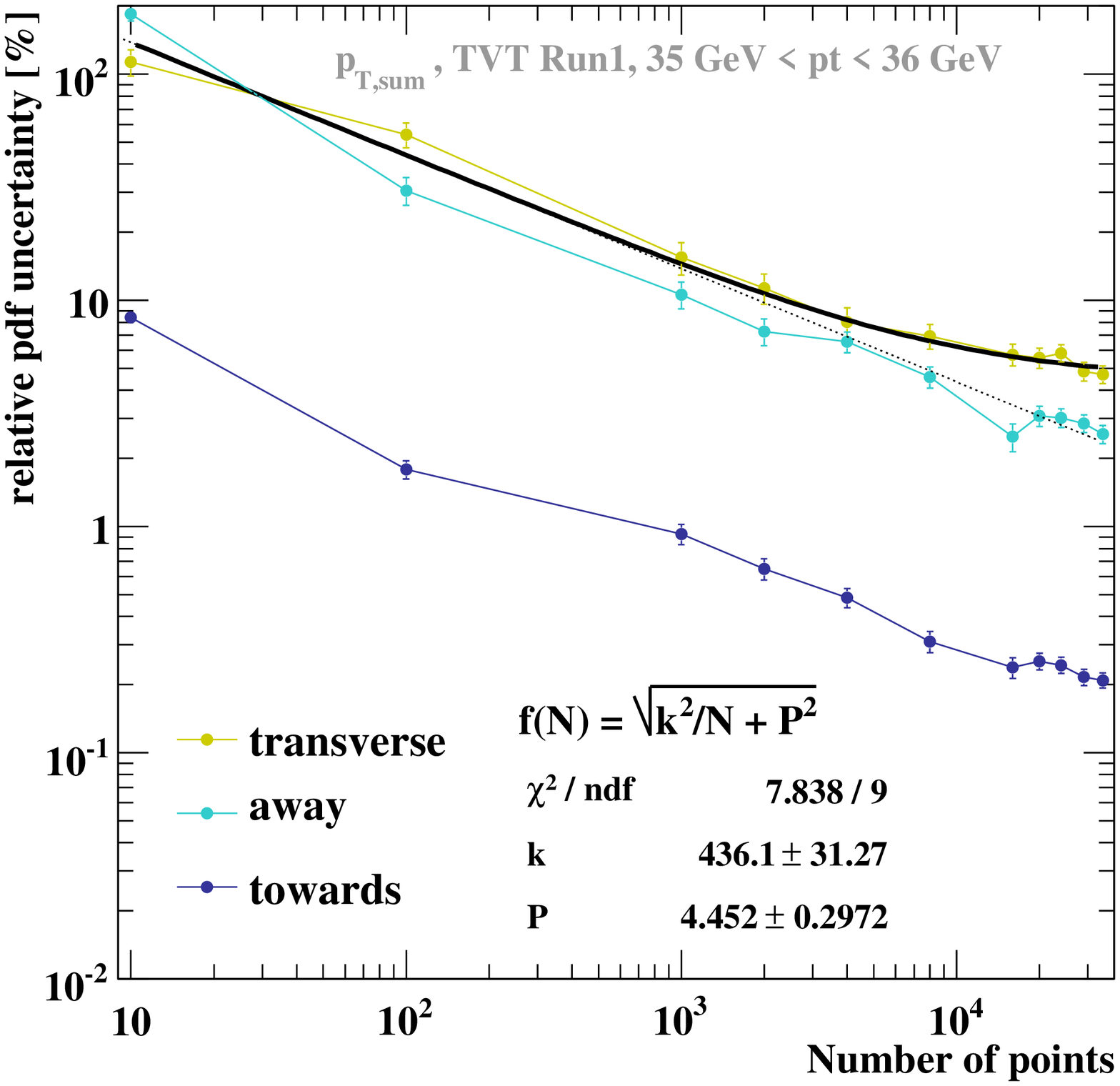}
  \caption{
    \label{fig:pdferror}
    Relative PDF uncertainty, $\Delta X/X(S_0)$, in percent. Left for
    the multiplicity observables and right for the $p_T^{\rm sum}$
    observables. The different curves show the results for the three
    different regions defined in the experimental analysis. The PDFs
    used are CTEQ6M \cite{Pumplin:2002vw} and its corresponding error
    sets. The fit result shown as a solid line is for the transverse
    region. Also shown as a light dashed line is the fit assuming a
    purely statistical error.}
}

As an example, we show the result in
Fig.~\ref{fig:pdferror} for one bin corresponding to $35-36$ GeV of the
leading jet. The final statistics are obtained from 20M fully
generated events for each PDF set and the value on the $x$ axis is the
number of events falling within this bin. We see that with these 20M
events, we have still not completely eliminated the statistical
uncertainties. However, a departure from the straight line on a
log--log plot that would be expected for pure statistical errors,
$\sim1/\sqrt{N}$, is clearly observed. We use this to extract the
\emph{true} PDF uncertainty, by fitting a curve of the form
\begin{equation}\label{eq:fit}
  f(N) = \sqrt{\frac{k^2}{N} + \mathrm{P}^2}
\end{equation}
to these data. In performing the fit we get a reliable result already
for a moderate number of events. From the fit results we can estimate
the number of events that would be necessary to eliminate the
contribution of the statistical uncertainty. Requiring it to be less
than $10^{-1}$ of the total uncertainty leads to $N \sim 10^{6}$,
which translates into $\sim 10^9 $ fully generated events for each of
the 40 PDF sets, which is not feasible in practice. Instead, using our
fit, we have a clear indication that the PDF uncertainty is around 4\%
for the multiplicity and 4.5\% for the $p_T^{sum}$ in the transverse
region.

It is note-worthy that the difference between the two PDF sets is larger
than the uncertainty on each. Although, as we have already mentioned,
the \ue{} will have already been measured before making precision
measurements or using jet cross sections to extract PDFs, a model tuned
to that \ue{} measurement will have to be used and its tuning will
depend on the PDF set. We consider an uncertainty of 5--10\% large
enough to warrant further study in this direction.

\subsection{LHC expectation}

We start the discussion of our predictions for the LHC with the plots in
Fig.~\ref{fig:obs_lhc}, which are related to the total multiplicity and
mean multiplicity flow in jet events. We show \HWPP\ with and without
MPI. We used QCD jet production with a minimal $p_T$ of 20~GeV as signal
process. The MPI parameters were left at their default values, i.e.\ the
fit to Tevatron CDF data.

The first plot in Fig.~\ref{fig:obs_lhc} shows the KNO
distribution\cite{Koba:1972ng}. The MPI model satisfies KNO scaling
fairly well, whereas \HWPP\ without an \ue{} clearly violates it.

The second plot in Fig.~\ref{fig:obs_lhc} shows the mean charged
multiplicity as a function of pseudorapidity, $\eta$.  The effect of MPI
is clearly
visible, growing significantly from the Tevatron to the LHC.

\FIGURE[t]{
  \includegraphics[%
    width=.48\textwidth,keepaspectratio]{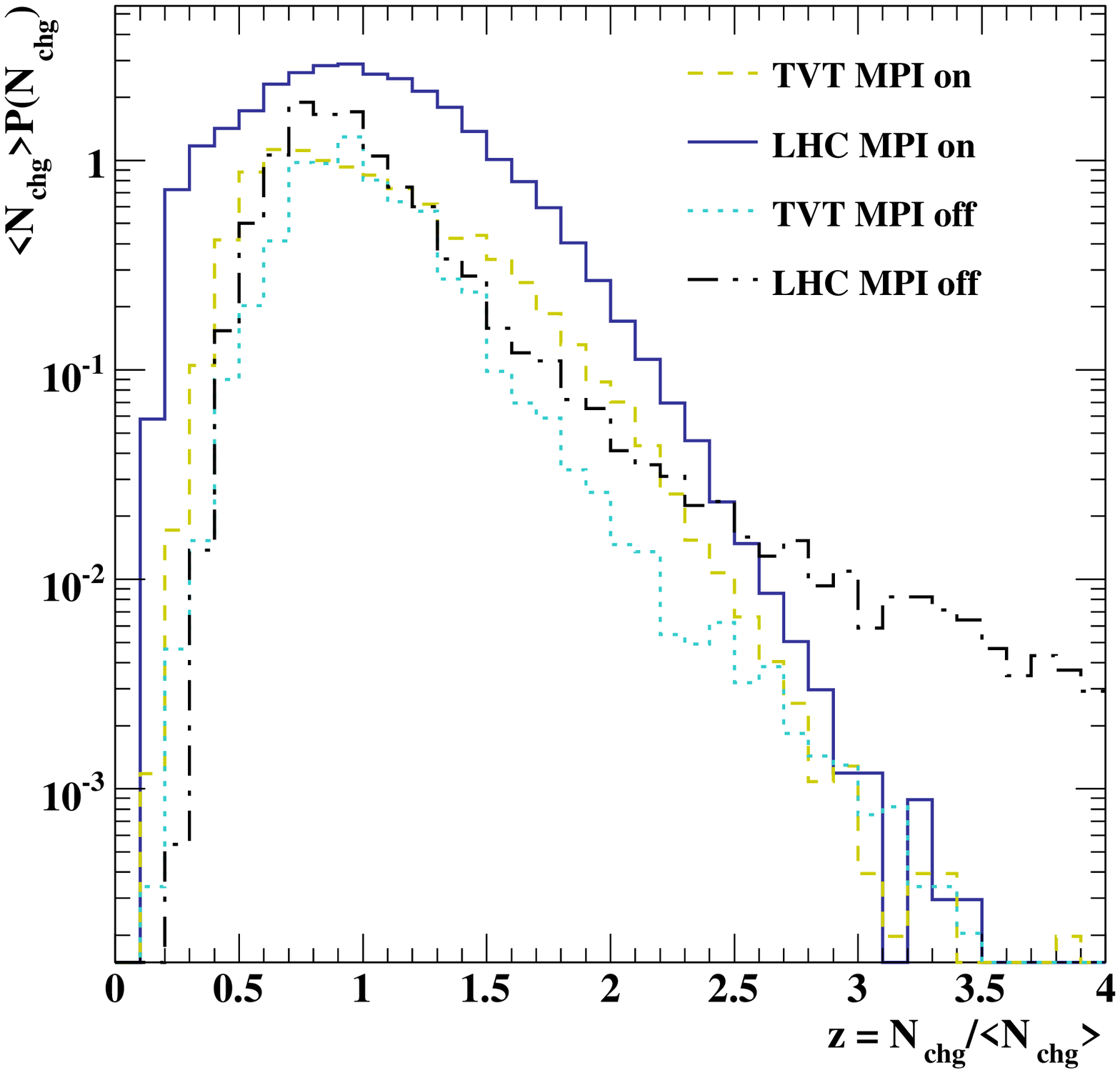}
  \includegraphics[%
    width=.48\textwidth,keepaspectratio]{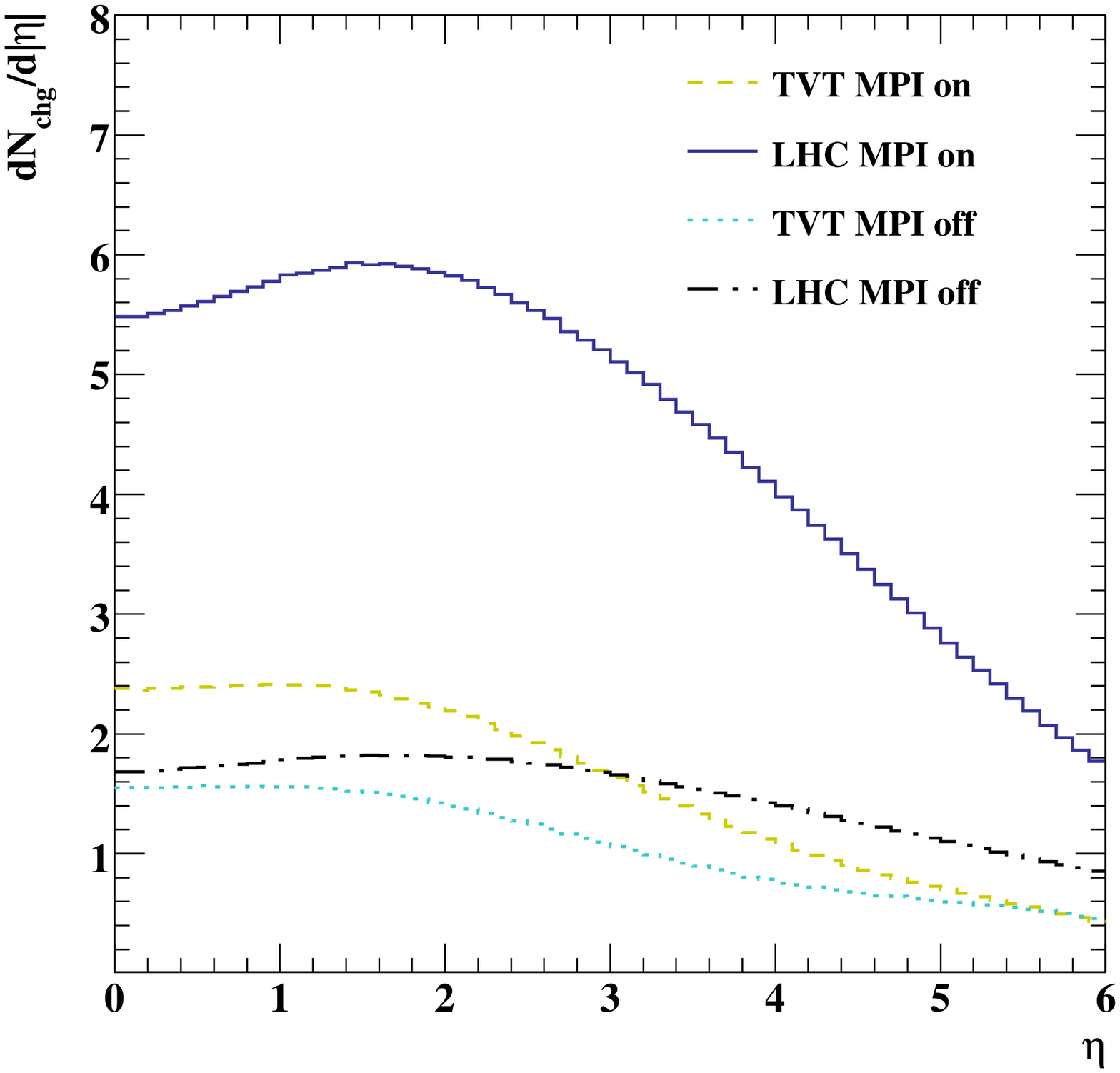}
  \caption{
    \label{fig:obs_lhc}
    KNO plot (left) and differential multiplicity distribution (right)
    for Tevatron and LHC runs.}
} 

In Ref.~\cite{Alekhin:2005dx} a comparison of different predictions for
an analysis modelled on the CDF one discussed earlier was presented. As
a benchmark observable the charged particle multiplicity for the
transverse region was used. All expectations reached a plateau in this
observable for $p_T^{ljet} > 10$~GeV. Our prediction for this observable
is shown in Fig.~\ref{fig:transv_lhc}, where it can be seen to have also
reached a constant plateau within the region shown.  The height of this
plateau can be used for comparison.  In Ref.~\cite{Alekhin:2005dx}
PYTHIA 6.214 ATLAS tune reached a height of $\sim 6.5$, PYTHIA 6.214
CDF Tune~A of $\sim 5$ and PHOJET 1.12 of $\sim 3$. Our model reaches
a height of $\sim 5$ and seems to be close to the PYTHIA 6.214 CDF
tune, although our model parameters were kept constant at their values
extracted from the fit to Tevatron data.

\FIGURE[t]{
  \includegraphics[%
    width=.48\textwidth,keepaspectratio]{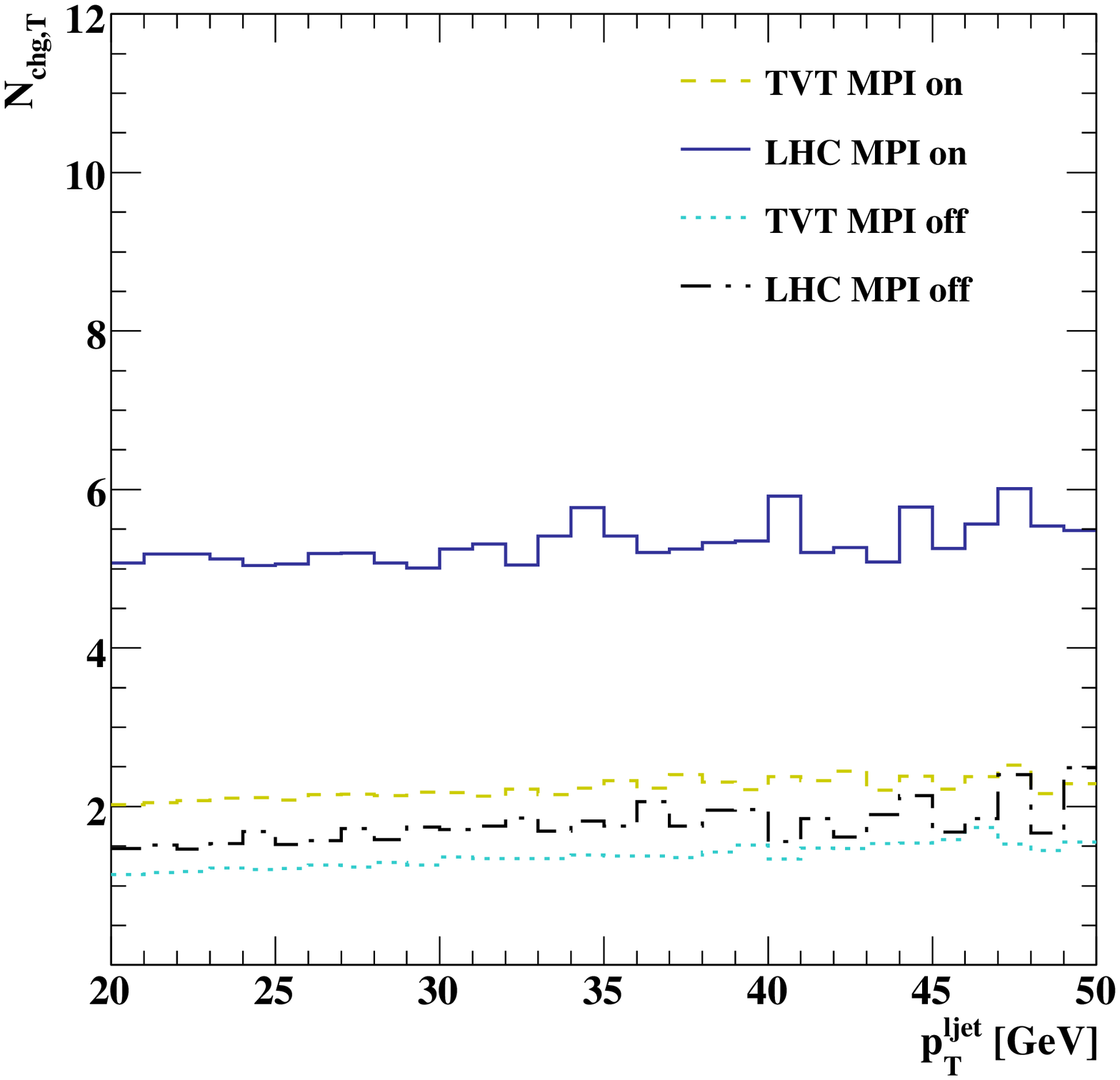}
  \includegraphics[%
    width=.48\textwidth,keepaspectratio]{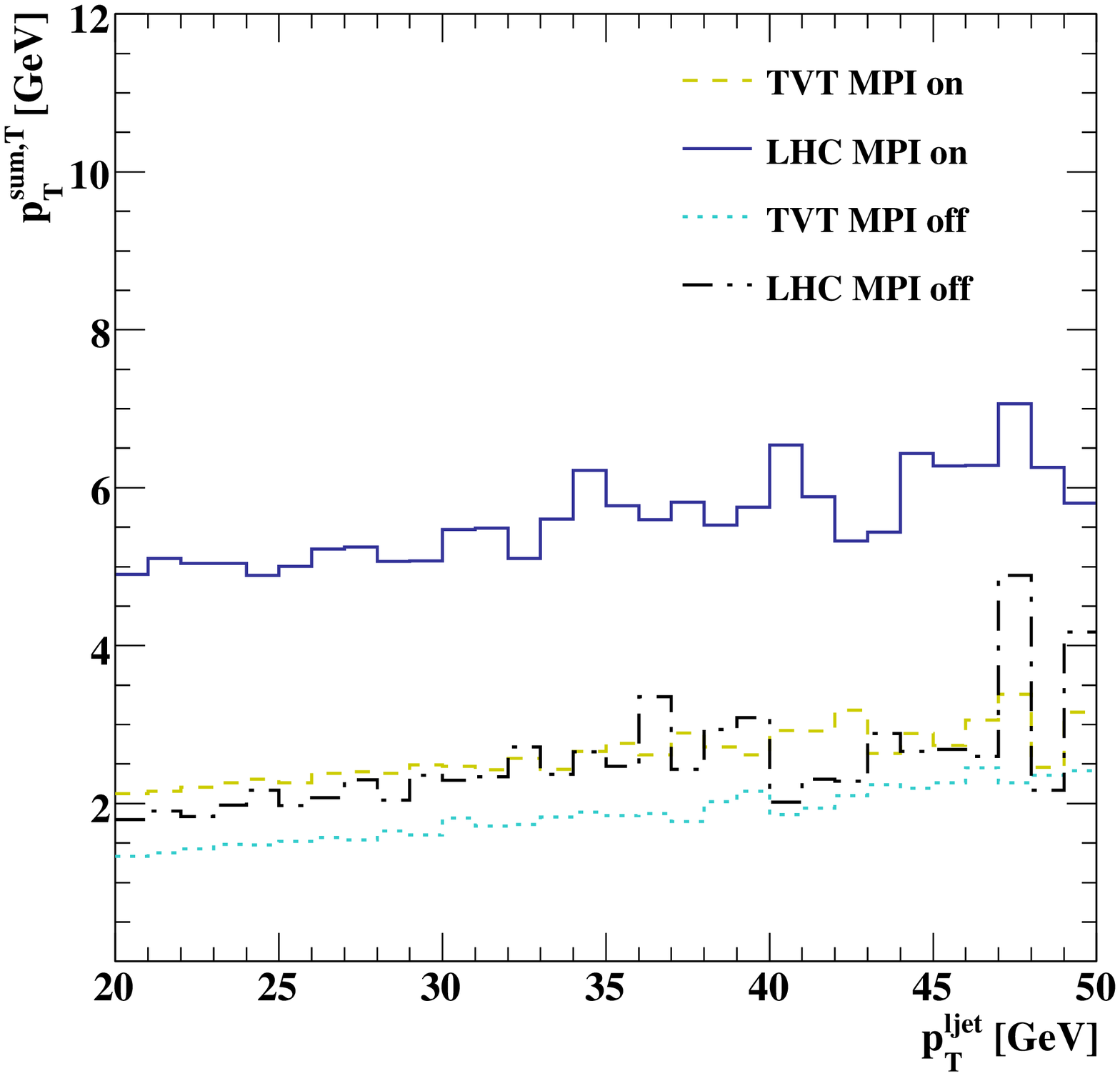}
  \caption{
    \label{fig:transv_lhc}
    Multiplicity and ptsum in the transverse region for LHC runs with
    \HWPP. The different data sets are (from bottom to top): Tevatron
    with MPI off, LHC with MPI off, Tevatron with MPI on and LHC with
    MPI on.}                                    
} 

We have seen already in Sec.~\ref{sec:TVTtune} that our fit results in a
flat valley of parameter points, which all give a very good description
of the data. We will briefly estimate the spread of our LHC
expectations, using only parameter sets from this
valley. The range of predictions that we deduce will be the range that
can be expected assuming no energy dependence on our main
parameters. Therefore early measurements could shed light into the
potential energy dependence of the input parameters by simply comparing
first data to these predictions. We extracted the average value of the
two transverse observables shown in Fig.~\ref{fig:transv_lhc} for a
given parameter set in the region $20 \GeV < p_T^{ljet} < 30 \GeV$. We
did that for the best fit points at three different values for $\ptmin$,
namely 2 GeV, 3.4 GeV and 4.5 GeV.

\renewcommand{\arraystretch}{1.2}

\begin{center}
  \begin{tabular}{l|c|c}
    LHC predictions &$\langle N_{chg}\rangle^{transv}$ & $\langle
    p_T^{sum}\rangle^{transv} [\GeV]$\\
    \hline
    TVT best fit& $5.1 \pm 0.3$ & $5.0 \pm 0.5$\\
  \end{tabular}
\end{center}

\section{Conclusions\label{sec:Conclusions}}

We have implemented a model of multiple parton interactions into the
\hw{} event generator.  We have tuned its free parameters to Tevatron
data and found a good overall description.  We have shown the
extrapolation of its predictions to the LHC.

We consider the present work as only a first step towards our eventual
goal of providing a complete description of the final state of minimum
bias collisions and \ue{}s in hard hadron--hadron collisions, validated on
and tuned to all available data and extrapolated to the LHC with
quantified uncertainties.

Among the various phenomenological and theoretical studies that will be
needed to achieve this goal, we mention the following avenues for future
work.  The present model only considers the contribution to multiple
scattering from perturbative processes above $\ptmin$.  We plan to
extend it along the lines discussed in Ref.~\cite{Borozan:2002fk} to
include non-perturbative partonic scattering below $\ptmin$.  This will
allow a description of minimum bias events, as well as the \ue{}.  There
is a lot more data available that constrains \ue{} and minimum bias
models to varying degrees.  We plan to make a global analysis of this,
in particular to give a handle on the energy dependence.  It would also
be interesting to consider whether the data require or allow an energy-
(and scale-) -dependent effective proton radius, as predicted in
Ref.\cite{Godbole:2004kx}.  Finally, it would be interesting to explore
whether saturation effects are important and whether multiparton
correlations, as discussed in \cite{Rogers:2008ua}, can be incorporated.

\section*{Acknowledgements}

We would like to thank our collaborators on the \hw\ project for many
useful discussions. Inspiring discussions with Jon Butterworth and the
indispensible help of Leif L\"onnblad are gratefully acknowledged. MB
thanks the hospitality of the HEP group of UCL, where this study was
completed. This work was supported in part by the European Union Marie
Curie Research Training Network MCnet under contract MRTN-CT-2006-035606
and the Helmholtz--Alliance ``Physics at the Terascale''.  MB was
supported by the Landesgraduiertenf\"orderung Baden-W\"urttemberg.

\appendix

\section{Forced splitting: implementation in the event record}
\label{appA}

\FIGURE[h]{
  \includegraphics[%
    width=.68\textwidth,keepaspectratio]{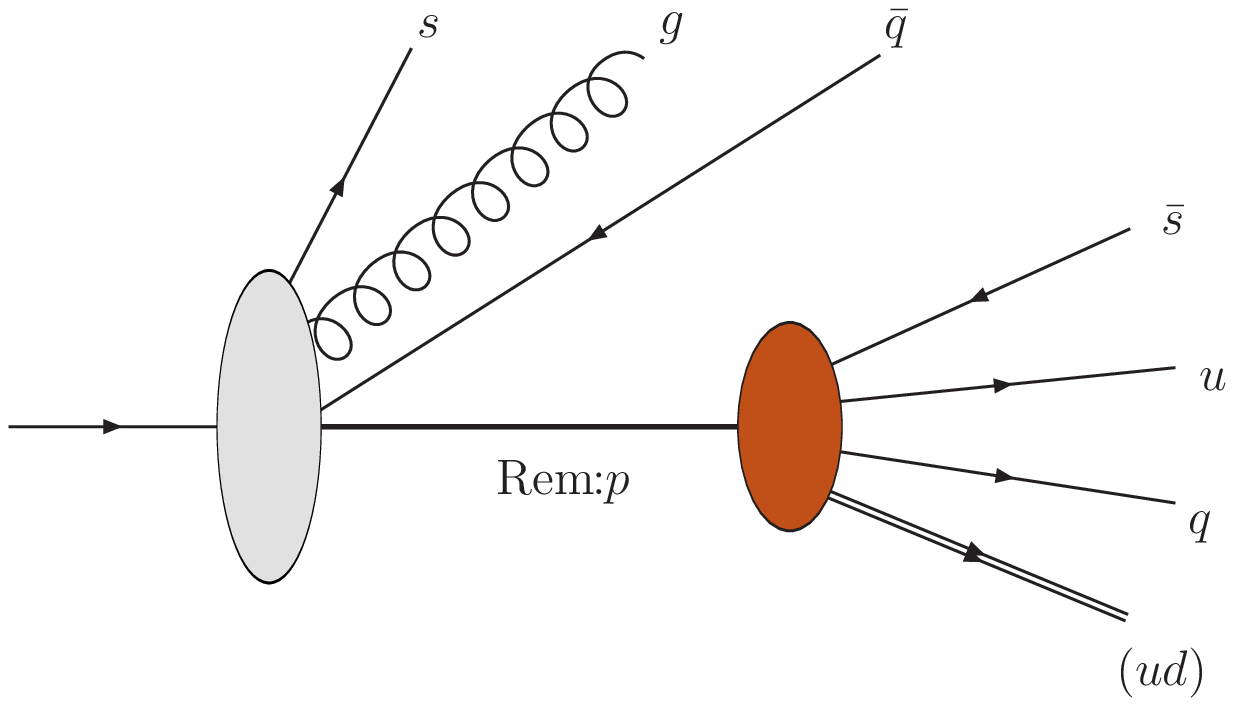}
  \caption{
    \label{fig:schema2}
    The structure of the event as it is implemented in \hw{}.}
} 

In Sect.~\ref{sec:details} we have briefly described how the different
hard scatters are correlated in colour space.  This is of course an
important model detail.  In the event record, however, this will not be
very obvious as this appears to be organized differently, in a way more
closely related to the eikonal idea.  In Fig.~\ref{fig:schema2} we show
the same particles ($s, g, \bar q$) that have already been extracted
from the proton in the example of Fig.~\ref{fig:schema}.  This time the
particles that have been extracted as last particles of the parton
shower are directly extracted from the proton.  All additional emissions
of partons that are related to the forced splitting, described in
Sect.~\ref{sec:details} appear as decay products of the intermediate
remnant.  In this way we emphasize the non--perturbative origin of these
partons and draw a clear line between the perturbative parton shower
model and the non--perturbative mechanism of forced splittings in the
event record.

\section{Model parameters}
\label{appB}
In the \HWPP\ manual\cite{Bahr:2008pv}, the general mechanism for
accessing and changing parameters and switches of models is described,
together with the main parameters and switches of the \ue\ model.  For
completeness, we repeat the latter here.

\subparagraph{\protect\HWPPParameter{MPIHandler}{Cuts}:}
   Via a cuts object the minimal $p_T$ of the additional scatters can be
   set. This is one of the two main parameters of the model. The current
   default, obtained from the fit to Tevatron data described above, is
   $3.4\, \mathrm{GeV}$.
\subparagraph{\protect\HWPPParameter{MPIHandler}{InvRadius}:} 
   The inverse beam particle radius squared. The current default is
   $1.5\, \mathrm{GeV}^{2}$, obtained from the above mentioned fit.
\subparagraph{\protect\HWPPParameter{MPIHandler}{Algorithm}:} 
   A switch to enable efficient generation of additional scatters in
   rare (high-$p_T$) signal processes.  Steers whether to use
   Eq.~(\ref{eq:prob0}) or Eq.~(\ref{eq:prob1}). The options are:
   \begin{itemize}
     \item 0: \Ue\ process and signal process are identical.
     \item 1: \Ue\ process and signal process are of the same type but
       the signal cross section is small. Here a veto
       algorithm has to be applied, if an additional scatter is produced
       with $p_T$ larger than the cutoff for the hard process.
     \item 2: \Ue\ process and signal process are distinct scattering
       types and the signal cross section is small. This is the default
       choice.
   \end{itemize}

\section{\boldmath Systematic errors in the low $p_T$ region}\label{app:syserr}

When making the initial comparison with data, we observed a $>3\sigma$
discrepancy for the observable $p_{T,sum}^{tow}$ below 30 GeV of the
leading jet. Above 30 GeV, this discrepancy is completely
absent. However, we have almost no freedom to tune this observable,
because it is completely dominated by the $p_T$ of the jet itself. For
the same reason, the relative error is extremely small in this region,
$\sim0.5\%$, so the absolute discrepancy is only about 2\%. Nevertheless
if we are going to fit to data in this region, we need to understand
this effect, to avoid the $\chi^2$ of the fit being completely
dominated by it.

From Ref.~\cite{Affolder:2001xt} we find that the data sample was
obtained by requiring a calorimeter tower with $E_T > 20 \GeV$
(including charged and neutral particles), described as the `Jet20'
sample. The analysis however is based
on charged particle tracks. In particular the $x$-axis in all observables
is the scalar $p_T$ sum of the charged particles defined to be in the
hardest jet. It is clear that this sample is only unbiased for large enough
values of $p_T^{ljet}$ relative to the 20~GeV trigger.
Where this happens however is not
obvious. In Ref.~\cite{Affolder:2001xt} the sample was assumed to be
perfectly unbiased from 20~GeV onwards. This statement is based on the
good match between the Jet20 data and the Min Bias sample around that
value. Any judgement on the smoothness of the match is however limited
by the statistical error on the Min Bias data, which is becoming large
in the region where the two samples overlap. Therefore we have added an
additional systematic error in quadrature to the data points to reflect
the precision with which we are confident they are unbiased. We choose
this to have the form
\begin{equation}
  \sigma_{add} = \frac{\sigma_{sys}^0}{10} \ \left( 30.5 -
  \frac{p_T}{\GeV} \right) \text{ for } (20.5 < p_T/\GeV < 30.5) \ ,
\end{equation}
where $\sigma_{sys}^0$ is extracted from the uncertainties in
the bins $18-21 \GeV$ of the Min Bias data and the linear form ensures
that the additional uncertainty goes to zero for $p_t\sim30$~GeV. In
more detail, we extract $\sigma_{sys}^0$ by fitting these three bins
with a linear function and use the uncertainty on the value at 20.5~GeV
from this fit for $\sigma_{sys}^0$ (in practice, this procedure gives
only a slightly smaller error than simply averaging the errors on the
three bins). This gives the following values for the $p_{T,sum}$
observables:
\begin{center}
  \begin{tabular}{c|c}
    region & $\sigma_{sys}^0$ \\
    \hline
    towards    & 440  MeV\\
    away       & 1950 MeV\\
    transverse & 840  MeV
  \end{tabular}
\end{center}
For the multiplicities we obtain the following values:
\begin{center}
  \begin{tabular}{c|c}
    region & $\sigma_{sys}^0$ \\
    \hline
    towards    & 0.75  \\
    away       & 1.07  \\
    transverse & 0.63 
  \end{tabular}
\end{center}

\bibliographystyle{JHEP}
\bibliography{UEpaper.bib}

\providecommand{\href}[2]{#2}\begingroup\raggedright\begin{thebibliography}{10}

\bibitem{Corcella:2000bw}
G.~Corcella {\em et.~al.}, {\it {HERWIG 6: An event generator for hadron
  emission reactions with interfering gluons (including supersymmetric
  processes)}},  {\em JHEP} {\bf 01} (2001) 010,
  [\href{http://xxx.lanl.gov/abs/hep-ph/0011363}{{\tt hep-ph/0011363}}].

\bibitem{Gieseke:2003hm}
S.~Gieseke, A.~Ribon, M.~H. Seymour, P.~Stephens, and B.~Webber, {\it {Herwig++
  1.0: An event generator for $e^+e^-$ annihilation}},  {\em JHEP} {\bf 02}
  (2004) 005, [\href{http://xxx.lanl.gov/abs/hep-ph/0311208}{{\tt
  hep-ph/0311208}}].

\bibitem{Gieseke:2006ga}
S.~Gieseke {\em et.~al.}, {\it Herwig++ 2.0 release note},
  \href{http://xxx.lanl.gov/abs/hep-ph/0609306}{{\tt hep-ph/0609306}}.

\bibitem{Bahr:2007ni}
M.~B\mbox{\"{a}}hr {\em et.~al.}, {\it Herwig++ 2.1 release note},
  \href{http://xxx.lanl.gov/abs/0711.3137}{{\tt arXiv:0711.3137}}.

\bibitem{Bahr:2008pv}
M.~B\mbox{\"{a}}hr {\em et.~al.}, {\it {Herwig++ Physics and Manual}},
  \href{http://xxx.lanl.gov/abs/0803.0883}{{\tt arXiv:0803.0883}}.

\bibitem{Bahr:2008tx}
M.~B\mbox{\"{a}}hr {\em et.~al.}, {\it Herwig++ 2.2 release note},
  \href{http://xxx.lanl.gov/abs/0804.3053}{{\tt arXiv:0804.3053}}.

\bibitem{Sjostrand:2006za}
T.~Sj\mbox{\"{o}}strand, S.~Mrenna, and P.~Skands, {\it {PYTHIA 6.4 physics and
  manual}},  {\em JHEP} {\bf 05} (2006) 026,
  [\href{http://xxx.lanl.gov/abs/hep-ph/0603175}{{\tt hep-ph/0603175}}].

\bibitem{Sjostrand:2007gs}
T.~Sj\mbox{\"{o}}strand, S.~Mrenna, and P.~Skands, {\it {A Brief Introduction
  to PYTHIA 8.1}},  \href{http://xxx.lanl.gov/abs/0710.3820}{{\tt 0710.3820}}.

\bibitem{Gleisberg:2003xi}
T.~Gleisberg {\em et.~al.}, {\it {SHERPA 1.$\alpha$, a proof-of-concept
  version}},  {\em JHEP} {\bf 02} (2004) 056,
  [\href{http://xxx.lanl.gov/abs/hep-ph/0311263}{{\tt hep-ph/0311263}}].

\bibitem{Catani:2001cc}
S.~Catani, F.~Krauss, R.~Kuhn, and B.~R. Webber, {\it {QCD matrix elements +
  parton showers}},  {\em JHEP} {\bf 11} (2001) 063,
  [\href{http://xxx.lanl.gov/abs/hep-ph/0109231}{{\tt hep-ph/0109231}}].

\bibitem{Alwall:2007fs}
J.~Alwall {\em et.~al.}, {\it {Comparative study of various algorithms for the
  merging of parton showers and matrix elements in hadronic collisions}},  {\em
  Eur. Phys. J.} {\bf C53} (2008) 473--500,
  [\href{http://xxx.lanl.gov/abs/0706.2569}{{\tt 0706.2569}}].

\bibitem{Alner:1986is}
{\bf UA5} Collaboration, G.~J. Alner {\em et.~al.}, {\it {The UA5 High-Energy
  $\bar{p}p$ Simulation Program}},  {\em Nucl. Phys.} {\bf B291} (1987) 445.

\bibitem{Sjostrand:1987su}
T.~Sj\mbox{\"{o}}strand and M.~van Zijl, {\it {A Multiple Interaction Model for
  the Event Structure in Hadron Collisions}},  {\em Phys. Rev.} {\bf D36}
  (1987) 2019.

\bibitem{Butterworth:1996zw}
J.~M. Butterworth, J.~R. Forshaw, and M.~H. Seymour, {\it Multiparton
  interactions in photoproduction at \mbox{HERA}},  {\em Z. Phys.} {\bf C72}
  (1996) 637--646, [\href{http://xxx.lanl.gov/abs/hep-ph/9601371}{{\tt
  hep-ph/9601371}}].

\bibitem{JimmyManual}
J.~M. Butterworth and M.~H. Seymour, ``Jimmy4 {M}anual.'' Downloadable under
  http://projects.hepforge.org/jimmy/.

\bibitem{Capella:1978ig}
A.~Capella, U.~Sukhatme, C.-I. Tan, and J.~Tran Thanh~Van, {\it {Jets in Small
  p(T) Hadronic Collisions, Universality of Quark Fragmentation, and Rising
  Rapidity Plateaus}},  {\em Phys. Lett.} {\bf B81} (1979) 68.

\bibitem{Capella:1980fv}
A.~Capella and J.~Tran Thanh~Van, {\it {A New Parton Model Description of Soft
  Hadron-Nucleus Collisions}},  {\em Phys. Lett.} {\bf B93} (1980) 146.

\bibitem{Capella:1981xr}
A.~Capella and J.~Tran Thanh~Van, {\it {Hadron - Nucleus Interactions and the
  Leading Particle Effect in a Dual Parton Model}},  {\em Z. Phys.} {\bf C10}
  (1981) 249--262.

\bibitem{Engel:1994vs}
R.~Engel, {\it {Photoproduction within the two component dual parton model. 1.
  Amplitudes and cross-sections}},  {\em Z. Phys.} {\bf C66} (1995) 203--214.

\bibitem{Borozan:2002fk}
I.~Borozan and M.~H. Seymour, {\it {An eikonal model for multiparticle
  production in hadron hadron interactions}},  {\em JHEP} {\bf 09} (2002) 015,
  [\href{http://xxx.lanl.gov/abs/hep-ph/0207283}{{\tt hep-ph/0207283}}].

\bibitem{Akesson:1986iv}
{\bf Axial Field Spectrometer} Collaboration, T.~Akesson {\em et.~al.}, {\it
  {Double parton scattering in $pp$ collisions at $\sqrt{S} = 63\,$GeV}},  {\em
  Z. Phys.} {\bf C34} (1987) 163.

\bibitem{Pumplin:1997ix}
J.~Pumplin, {\it {Hard underlying event correction to inclusive jet cross
  sections}},  {\em Phys. Rev.} {\bf D57} (1998) 5787--5792,
  [\href{http://xxx.lanl.gov/abs/hep-ph/9708464}{{\tt hep-ph/9708464}}].

\bibitem{Abe:1993rv}
{\bf CDF} Collaboration, F.~Abe {\em et.~al.}, {\it {Study of four jet events
  and evidence for double parton interactions in $p\bar{p}$ collisions at
  $\sqrt{s} = 1.8$ TeV}},  {\em Phys. Rev.} {\bf D47} (1993) 4857--4871.

\bibitem{Abazov:2002mr}
{\bf D0} Collaboration, V.~M. Abazov {\em et.~al.}, {\it {Multiple jet
  production at low transverse energies in $p\bar{p}$ collisions at $\sqrt{s} =
  1.8$ TeV}},  {\em Phys. Rev.} {\bf D67} (2003) 052001,
  [\href{http://xxx.lanl.gov/abs/hep-ex/0207046}{{\tt hep-ex/0207046}}].

\bibitem{Abe:1997xk}
{\bf CDF} Collaboration, F.~Abe {\em et.~al.}, {\it {Double parton scattering
  in $\bar{p}p$ collisions at $\sqrt{s} = 1.8 $TeV}},  {\em Phys. Rev.} {\bf
  D56} (1997) 3811--3832.

\bibitem{Affolder:2001xt}
{\bf CDF} Collaboration, A.~A. Affolder {\em et.~al.}, {\it Charged jet
  evolution and the underlying event in $p\bar{p}$ collisions at
  1.8\,\mbox{TeV}},  {\em Phys. Rev.} {\bf D65} (2002) 092002.

\bibitem{Acosta:2004wqa}
{\bf CDF} Collaboration, D.~E. Acosta {\em et.~al.}, {\it {The underlying event
  in hard interactions at the Tevatron $\bar{p}p$ collider}},  {\em Phys. Rev.}
  {\bf D70} (2004) 072002, [\href{http://xxx.lanl.gov/abs/hep-ex/0404004}{{\tt
  hep-ex/0404004}}].

\bibitem{Moraes:2007rq}
A.~Moraes, C.~Buttar, and I.~Dawson, {\it {Prediction for minimum bias and the
  underlying event at LHC energies}},  {\em Eur. Phys. J.} {\bf C50} (2007)
  435--466.

\bibitem{Fano:2007zz}
L.~Fano, {\it {Minimum bias and underlying event at CMS}},  {\em Acta Phys.
  Polon.} {\bf B38} (2007) 435--442.

\bibitem{Acosta:2006bp}
D.~Acosta {\em et.~al.}, {\it {The underlying event at the LHC}}, .
  CERN-CMS-NOTE-2006-067.

\bibitem{Borjanovic:2004ce}
I.~Borjanovic {\em et.~al.}, {\it {Investigation of top mass measurements with
  the ATLAS detector at LHC}},  {\em Eur. Phys. J.} {\bf C39S2} (2005) 63--90,
  [\href{http://xxx.lanl.gov/abs/hep-ex/0403021}{{\tt hep-ex/0403021}}].

\bibitem{ChristophThesis}
C.~Hackstein, ``{Simulation of final states in vector boson fusion}.'' Diploma
  thesis, Insitut f\"{u}r Theoretische Physik, Universit\mbox{\"{a}}t
  Karlsruhe, Nov.~2007.

\bibitem{Sjostrand:2004pf}
T.~Sj\mbox{\"{o}}strand and P.~Z. Skands, {\it {Multiple interactions and the
  structure of beam remnants}},  {\em JHEP} {\bf 03} (2004) 053,
  [\href{http://xxx.lanl.gov/abs/hep-ph/0402078}{{\tt hep-ph/0402078}}].

\bibitem{Sjostrand:2004ef}
T.~Sj\mbox{\"{o}}strand and P.~Z. Skands, {\it {Transverse-momentum-ordered
  showers and interleaved multiple interactions}},  {\em Eur. Phys. J.} {\bf
  C39} (2005) 129--154, [\href{http://xxx.lanl.gov/abs/hep-ph/0408302}{{\tt
  hep-ph/0408302}}].

\bibitem{Alekhin:2005dx}
S.~Alekhin {\em et.~al.}, {\it {HERA and the LHC - A workshop on the
  implications of HERA for LHC physics: Proceedings Part A}},
  \href{http://xxx.lanl.gov/abs/hep-ph/0601012}{{\tt hep-ph/0601012}}.

\bibitem{Catani:1990eg}
S.~Catani, M.~Ciafaloni, and F.~Hautmann, {\it {High-energy factorization and
  small x heavy flavor production}},  {\em Nucl. Phys.} {\bf B366} (1991)
  135--188.

\bibitem{Collins:1991ty}
J.~C. Collins and R.~K. Ellis, {\it {Heavy quark production in very high-energy
  hadron collisions}},  {\em Nucl. Phys.} {\bf B360} (1991) 3--30.

\bibitem{Levin:1991ya}
E.~M. Levin, M.~G. Ryskin, Y.~M. Shabelski, and A.~G. Shuvaev, {\it {Heavy
  quark production in parton model and in QCD}},  {\em Sov. J. Nucl. Phys.}
  {\bf 54} (1991) 867--871.

\bibitem{Hoche:2007hg}
S.~H\mbox{\"{o}}che, F.~Krauss, and T.~Teubner, {\it {Multijet events in the
  k(T)-factorisation scheme}},  \href{http://xxx.lanl.gov/abs/0705.4577}{{\tt
  0705.4577}}.

\bibitem{Bartels:2005wa}
J.~Bartels, M.~Salvadore, and G.~P. Vacca, {\it {AGK cutting rules and multiple
  scattering in hadronic collisions}},  {\em Eur. Phys. J.} {\bf C42} (2005)
  53--71, [\href{http://xxx.lanl.gov/abs/hep-ph/0503049}{{\tt
  hep-ph/0503049}}].

\bibitem{Durand:1987yv}
L.~Durand and P.~Hong, {\it {QCD and Rising Cross Sections}},  {\em Phys. Rev.
  Lett.} {\bf 58} (1987) 303.

\bibitem{Durand:1988ax}
L.~Durand and H.~Pi, {\it {Semihard QCD and high-energy $pp$ and $\bar pp$
  scattering}},  {\em Phys. Rev.} {\bf D40} (1989) 1436.

\bibitem{Gieseke:2003rz}
S.~Gieseke, P.~Stephens, and B.~Webber, {\it {N}ew {F}ormalism for {QCD}
  {P}arton {S}howers},  {\em JHEP} {\bf 12} (2003) 045,
  [\href{http://xxx.lanl.gov/abs/hep-ph/0310083}{{\tt hep-ph/0310083}}].

\bibitem{Webber:1983if}
B.~R. Webber, {\it {A} {QCD} {M}odel for {J}et {F}ragmentation including {S}oft
  {G}luon {I}nterference},  {\em Nucl. Phys.} {\bf B238} (1984) 492.

\bibitem{Marchesini:1984bm}
G.~Marchesini and B.~R. Webber, {\it {S}imulation of {QCD} {J}ets including
  {S}oft {G}luon {I}nterference},  {\em Nucl. Phys.} {\bf B238} (1984) 1.

\bibitem{Marchesini:1987cf}
G.~Marchesini and B.~R. Webber, {\it {M}onte {C}arlo {S}imulation of {G}eneral
  {H}ard {P}rocesses with {C}oherent {QCD} radiation},  {\em Nucl. Phys.} {\bf
  B310} (1988) 461.

\bibitem{Grellscheid:2007tt}
D.~Grellscheid and P.~Richardson, {\it {S}imulation of {T}au {D}ecays in the
  {H}erwig++ {E}vent {G}enerator},
  \href{http://xxx.lanl.gov/abs/0710.1951}{{\tt 0710.1951}}.

\bibitem{MesonDecays}
D.~Grellscheid, K.~Hamilton, and P.~Richardson, ``{S}imulation of {M}eson
  {D}ecays in the {H}erwig++ {E}vent {G}enerator.'' in preparation.

\bibitem{Martin:2001es}
A.~D. Martin, R.~G. Roberts, W.~J. Stirling, and R.~S. Thorne, {\it {MRST2001:
  Partons and alpha(s) from precise deep inelastic scattering and Tevatron jet
  data}},  {\em Eur. Phys. J.} {\bf C23} (2002) 73--87,
  [\href{http://xxx.lanl.gov/abs/hep-ph/0110215}{{\tt hep-ph/0110215}}].

\bibitem{Gieseke:2004tc}
S.~Gieseke, {\it {Uncertainties of Sudakov form factors}},  {\em JHEP} {\bf 01}
  (2005) 058, [\href{http://xxx.lanl.gov/abs/hep-ph/0412342}{{\tt
  hep-ph/0412342}}].

\bibitem{Pumplin:2002vw}
J.~Pumplin {\em et.~al.}, {\it {New generation of parton distributions with
  uncertainties from global QCD analysis}},  {\em JHEP} {\bf 07} (2002) 012,
  [\href{http://xxx.lanl.gov/abs/hep-ph/0201195}{{\tt hep-ph/0201195}}].

\bibitem{Koba:1972ng}
Z.~Koba, H.~B. Nielsen, and P.~Olesen, {\it {Scaling of multiplicity
  distributions in high-energy hadron collisions}},  {\em Nucl. Phys.} {\bf
  B40} (1972) 317--334.

\bibitem{Godbole:2004kx}
R.~M. Godbole, A.~Grau, G.~Pancheri, and Y.~N. Srivastava, {\it {Soft gluon
  radiation and energy dependence of total hadronic cross-sections}},  {\em
  Phys. Rev.} {\bf D72} (2005) 076001,
  [\href{http://xxx.lanl.gov/abs/hep-ph/0408355}{{\tt hep-ph/0408355}}].

\bibitem{Rogers:2008ua}
T.~C. Rogers, A.~M. Stasto, and M.~I. Strikman, {\it {Unitarity Constraints on
  Semi-hard Jet Production in Impact Parameter Space}},
  \href{http://xxx.lanl.gov/abs/0801.0303}{{\tt 0801.0303}}.

\end{thebibliography}\endgroup

\FIGURE[htb]{
  \includegraphics[%
    width=\textwidth,keepaspectratio]{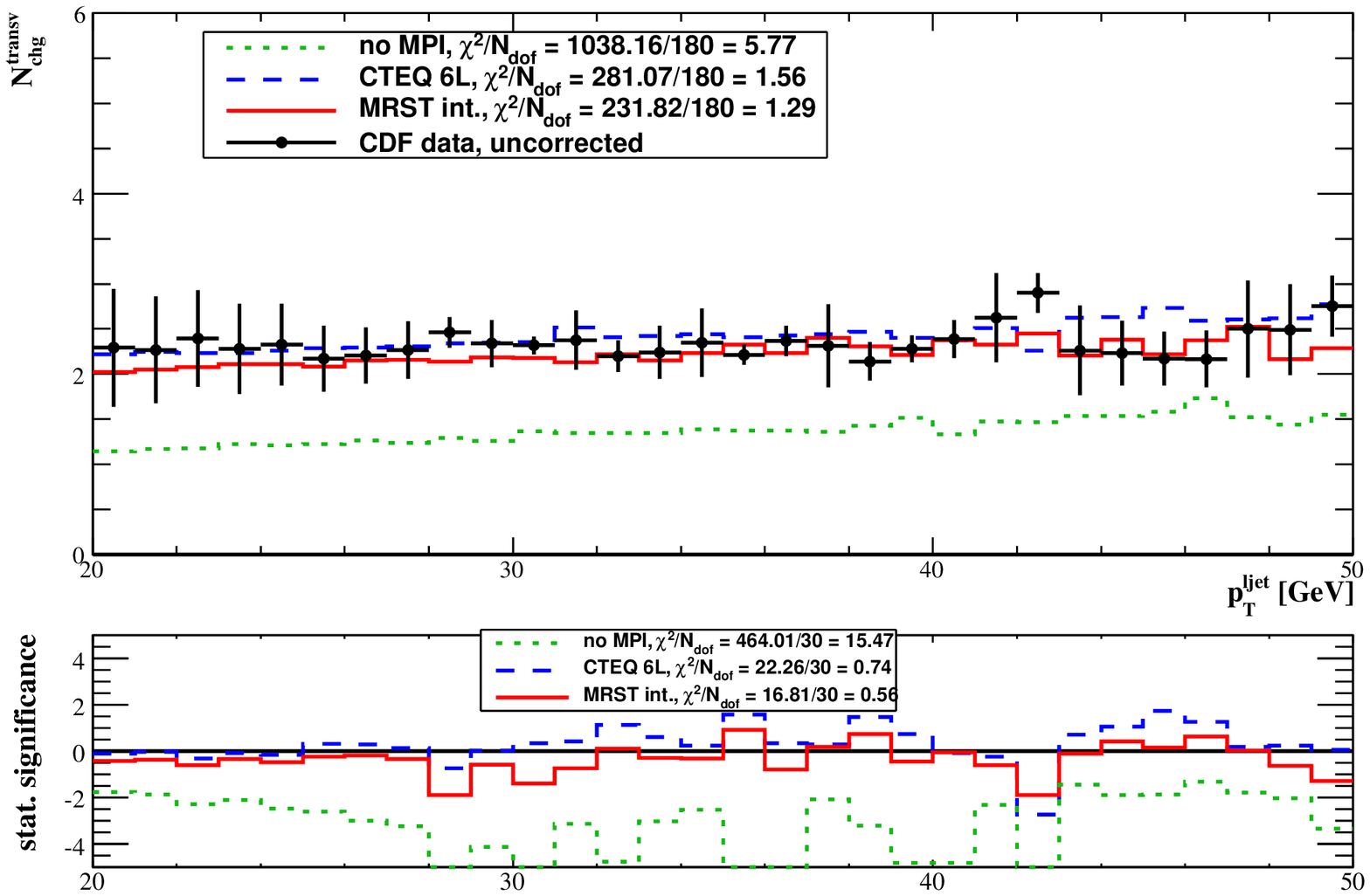}
  \\[0.4cm]
  \includegraphics[%
    width=\textwidth,keepaspectratio]{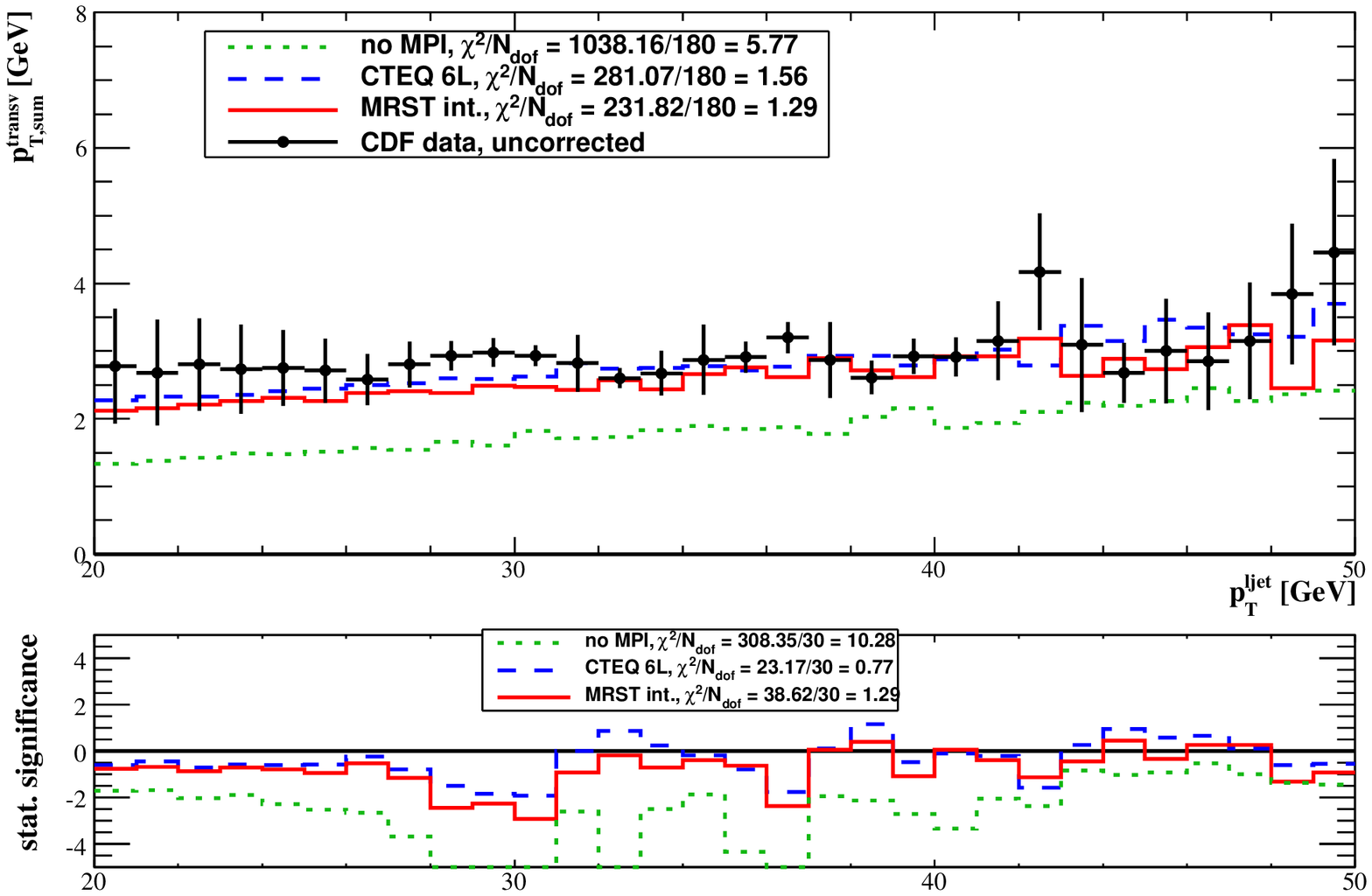}
  \caption{
    \label{fig:transv}
    Multiplicity and ptsum in the \textbf{transverse} region. CDF data
    are shown as black circles. \HWPP~ without MPI is drawn in green
    dots, \HWPP\ with MPI using MRST \cite{Martin:2001es} PDFs in solid
    red and with CTEQ6L \cite{Pumplin:2002vw} as blue dashed. The lower
    plot shows the statistical significance of the disagreement between
    Monte Carlo prediction and the data. The legend on the upper plot
    shows the total $\chi^2$ for all observables, whereas the lower plot
    has the $\chi^2$ values for this particular observable.}
} 

\FIGURE[htb]{
  \includegraphics[%
    width=\textwidth,keepaspectratio]{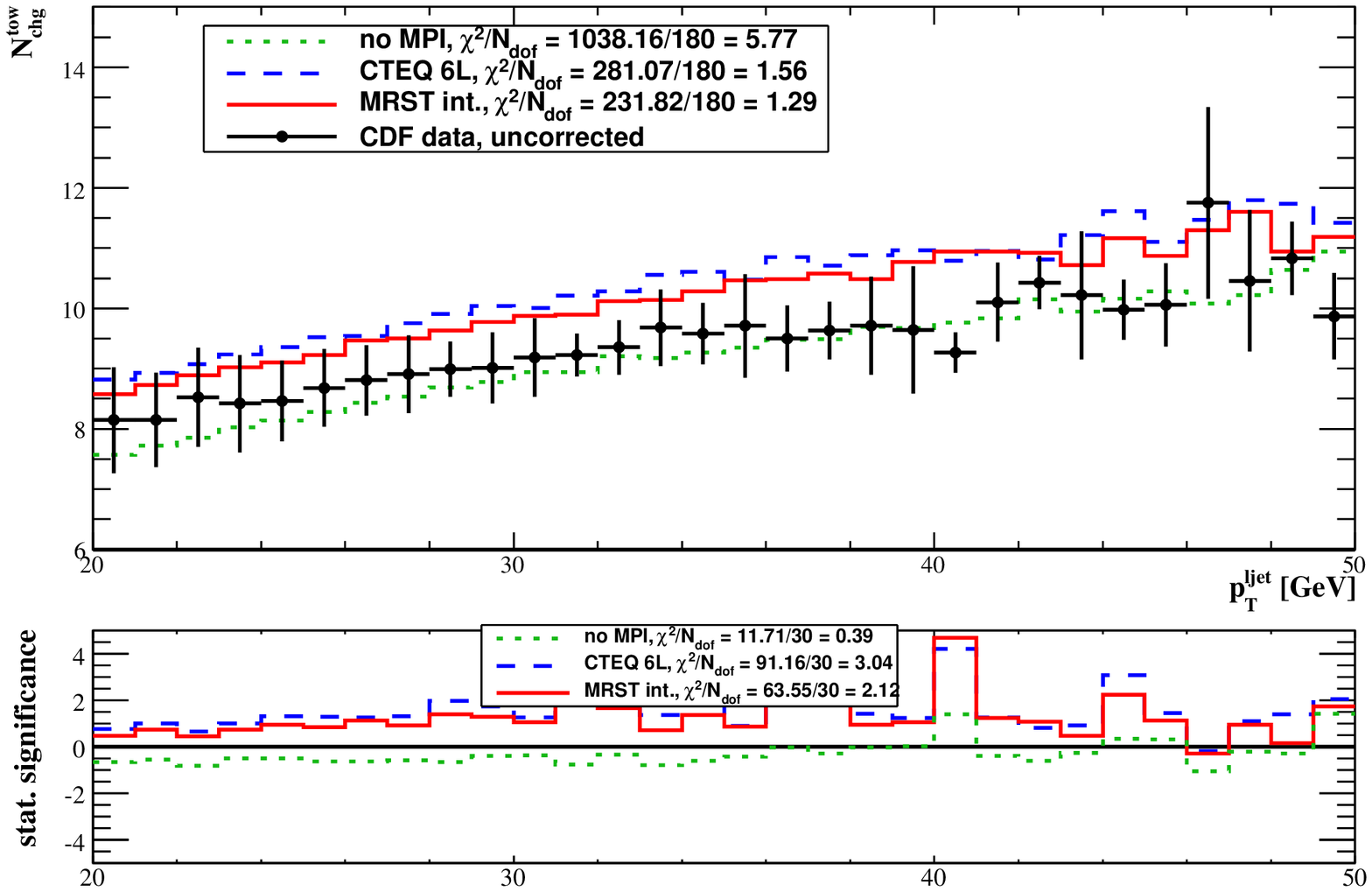}
  \\[0.4cm]
  \includegraphics[%
    width=\textwidth,keepaspectratio]{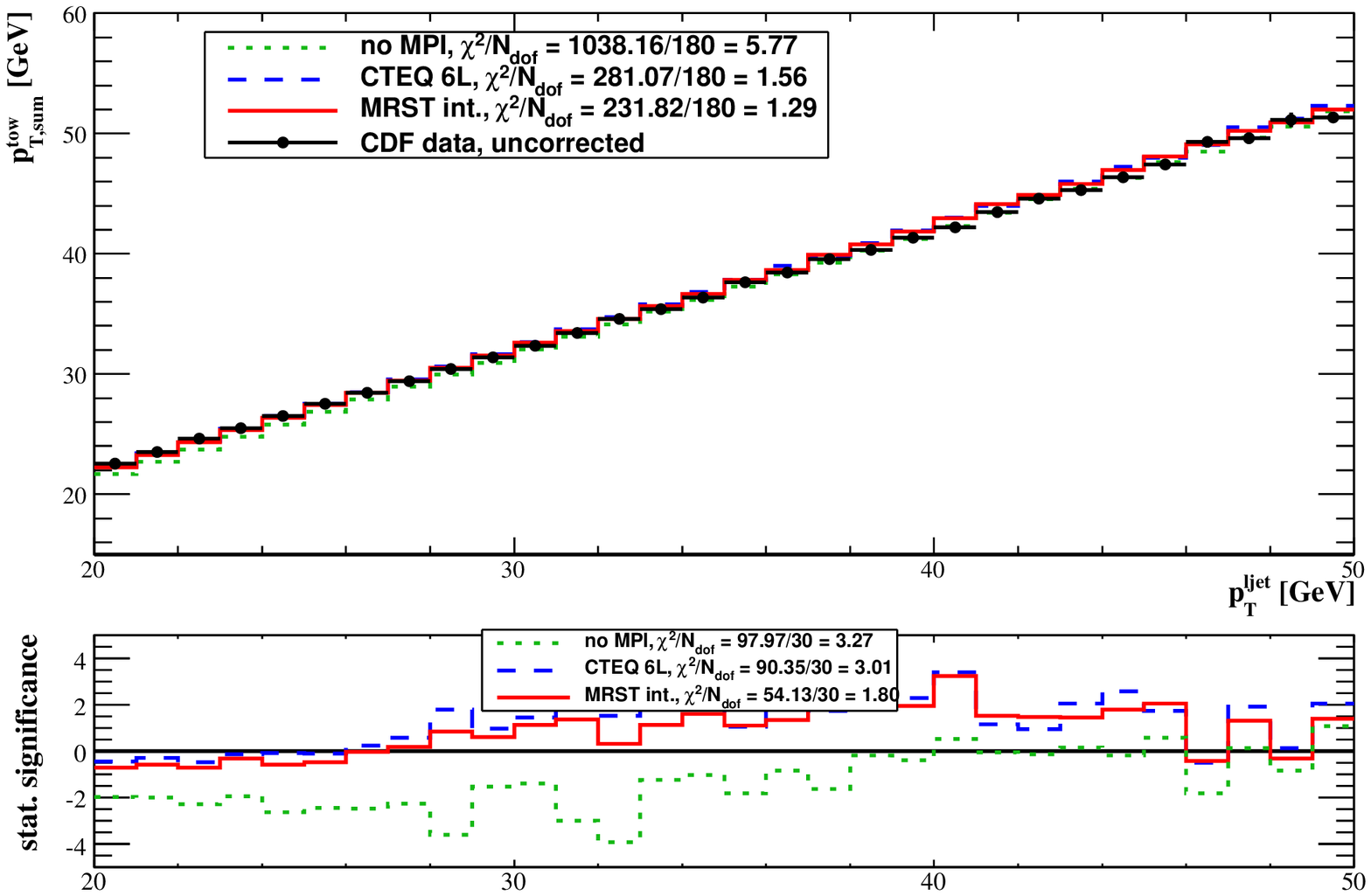}
  \caption{
    \label{fig:towards}
    Multiplicity and ptsum in the \textbf{towards} region. CDF data are
    shown as black circles. \HWPP~ without MPI is drawn in green dots,
    \HWPP\ with MPI using MRST \cite{Martin:2001es} PDFs in solid red
    and with CTEQ6L \cite{Pumplin:2002vw} as blue dashed. The lower plot
    shows the statistical significance of the disagreement between Monte
    Carlo prediction and the data. The legend on the upper plot shows
    the total $\chi^2$ for all observables, whereas the lower plot has
    the $\chi^2$ values for this particular observable.}
} 

\FIGURE[htb]{
  \includegraphics[%
    width=\textwidth,keepaspectratio]{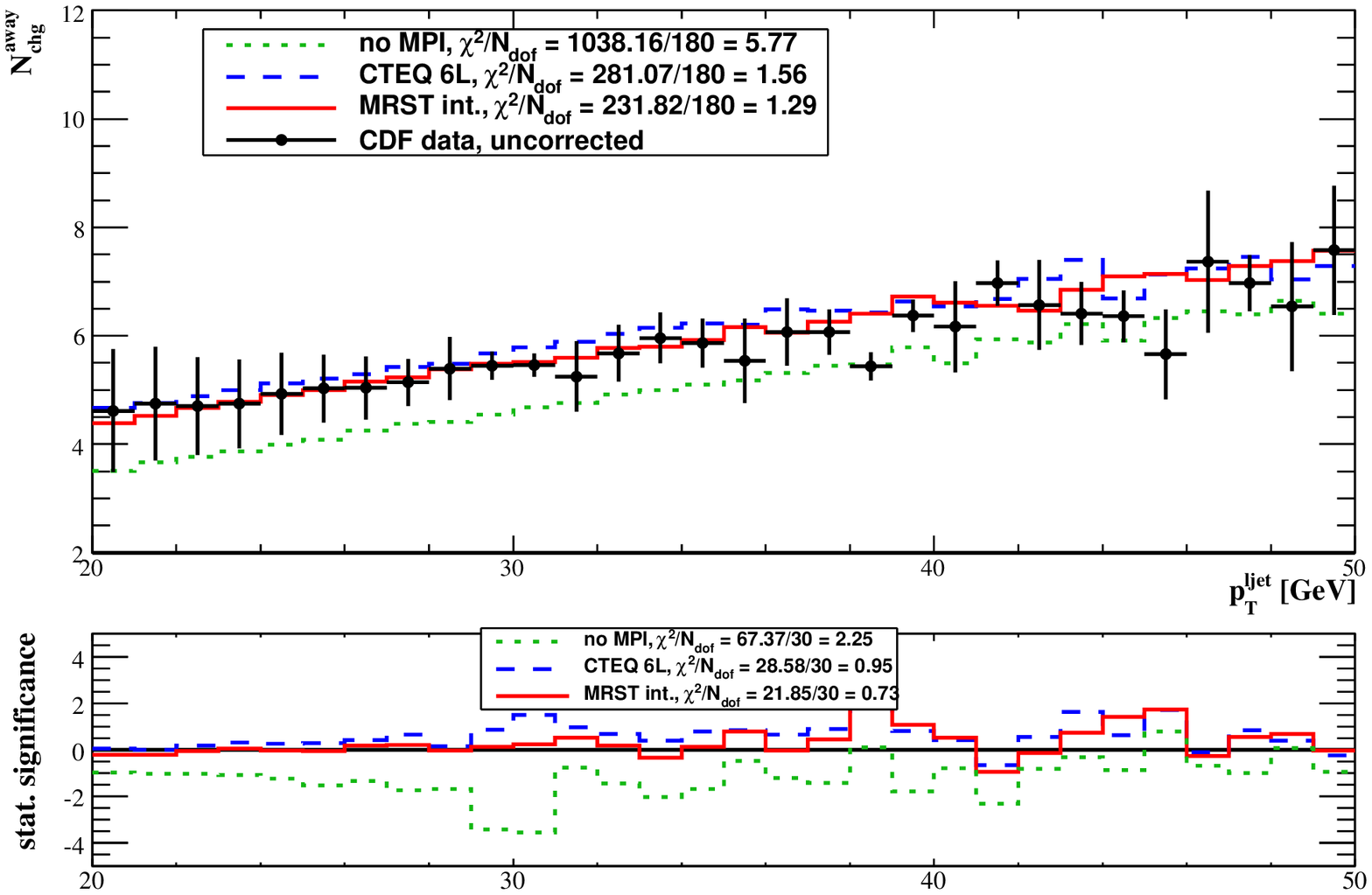}
  \\[0.4cm]
  \includegraphics[%
    width=\textwidth,keepaspectratio]{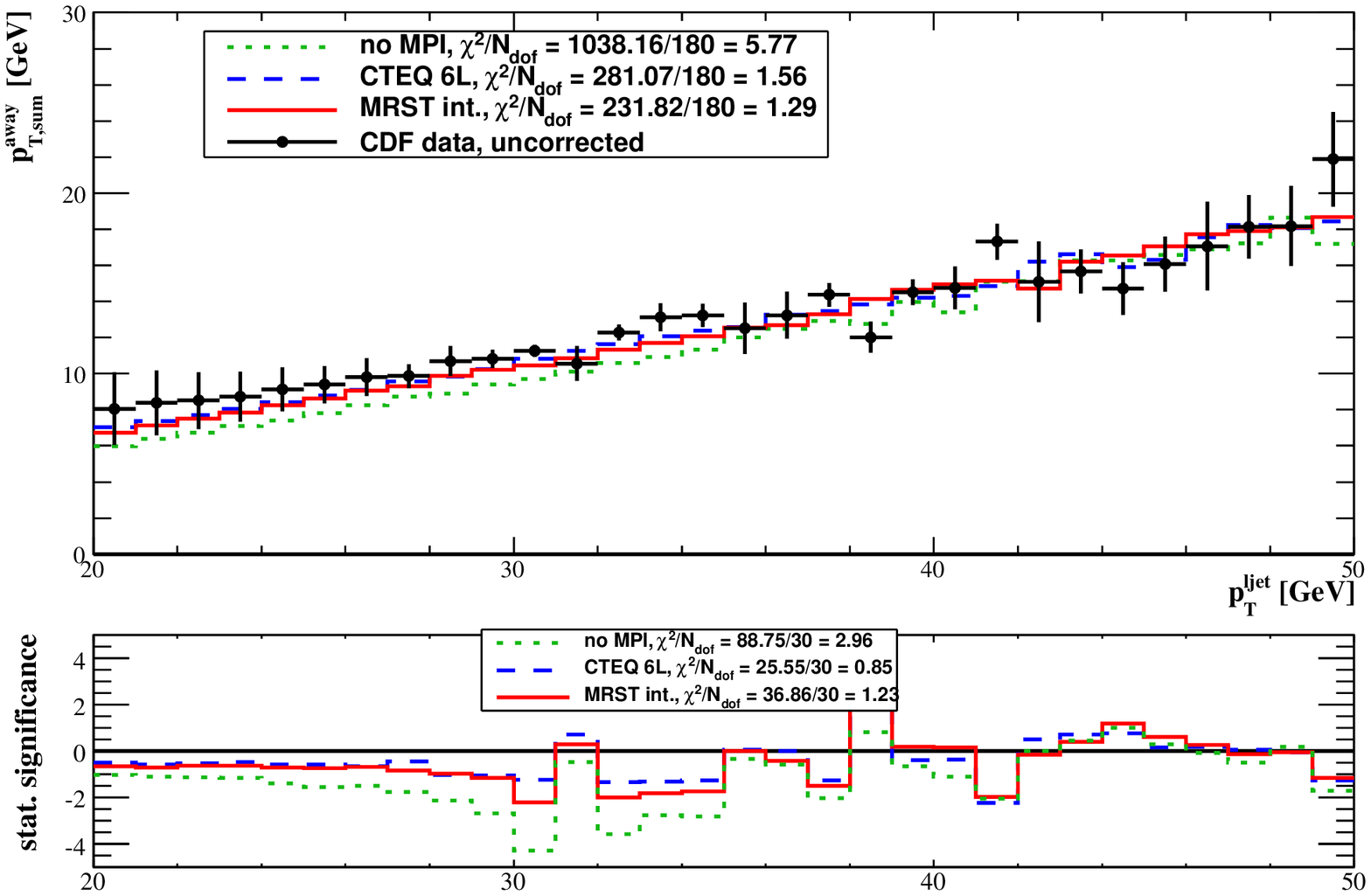}
  \caption{
    \label{fig:away}
    Multiplicity and ptsum in the \textbf{away} region. CDF data are
    shown as black circles. \HWPP~ without MPI is drawn in green dots,
    \HWPP\ with MPI using MRST \cite{Martin:2001es} PDFs in solid red
    and with CTEQ6L \cite{Pumplin:2002vw} as blue dashed. The lower plot
    shows the statistical significance of the disagreement between Monte
    Carlo prediction and the data. The legend on the upper plot shows
    the total $\chi^2$ for all observables, whereas the lower plot has
    the $\chi^2$ values for this particular observable.}
} 

\end{document}